\documentclass[%
 aip,
 amsmath,amssymb,
preprint,
]{revtex4-1}

\usepackage{graphicx}
\usepackage{dcolumn}
\usepackage{bm}

\usepackage[utf8]{inputenc}
\usepackage[T1]{fontenc}
\usepackage{mathptmx}
\usepackage{etoolbox}
\usepackage[caption=false]{subfig}

\makeatletter
\def\@email#1#2{%
 \endgroup
 \patchcmd{\titleblock@produce}
  {\frontmatter@RRAPformat}
  {\frontmatter@RRAPformat{\produce@RRAP{*#1\href{mailto:#2}{#2}}}\frontmatter@RRAPformat}
  {}{}
}%
\makeatother
\begin{document}

\title{Revisiting Rotary Blade Aerodynamics and Wake Behavior of a Vertical-Axis Wind Turbine via a High-Order Actuator-Line Framework}

\author{Abdullah Al Imran}
\thanks{Author to whom correspondence should be addressed}
 \email[Corresponding author: ]{aaimran84@umbc.edu.}
\email{aaimran84@umbc.edu, mlyu@umbc.edu}
\author{Meilin Yu}%
\affiliation{ 
University of Maryland, Baltimore County (UMBC), 1000 Hilltop Cir, Baltimore, MD, USA, 21250
}%

\date{\today}

\begin{abstract}
Vertical-axis wind turbines (VAWTs) experience highly unsteady blade loading and vortex shedding during operation, and the relationship between blade-wake interaction and useful power extraction remains unclear. The present study addresses this gap by relating phase-dependent blade loading and instantaneous wake activity to the mean velocity deficit and net power extraction across distinct operating regimes. Specifically, we show that deep dynamic stall produces large-scale vortex shedding and an energetic instantaneous wake, yet yields only a shallow mean velocity deficit and little net power. Near-optimal operation, by contrast, produces a more organized wake and substantially greater momentum extraction. To study the VAWT dynamics in a cost-effective way, we extended our previously developed steady source term formulation to a time-accurate, high-order numerical framework that couples flux reconstruction and actuator line model on fixed Cartesian grids, incorporating Gaussian force projection, a modified Boeing--Vertol dynamic stall model, and a double multiple streamtube based induction closure. The framework reproduces experimental near-wake profiles and predicts power coefficients comparable to those obtained from a three-dimensional large eddy simulation based actuator line model, with a peak power coefficient approximately 6\% lower than the reference value. An upwind-downwind force decomposition shows that dynamic stall acts predominantly during the upwind passage, temporarily increasing power under deep stall conditions while providing little net power overall. Light stall, by limiting large-scale flow separation and the subsequent loss of lift, maintains more consistent power-producing blade loading and therefore yields substantially greater mean power extraction.
\end{abstract}

\maketitle

\section{Introduction}
\label{sec:intro}
The growing demand for renewable energy has renewed interest in wind energy technologies that can operate under diverse environmental and installation conditions. Among these, vertical-axis wind turbines (VAWTs) offer geometric simplicity, omnidirectional operation, and potential advantages in turbulent and highly sheared inflow~\cite{paraschivoiu2002}. Their aerodynamic performance, however, remains difficult to predict because each blade experiences large cyclic variations in angle of attack, producing unsteady loading, dynamic stall, and repeated interactions with the rotor wake~\cite{simao2010simulating,lefouest2022dilemma}. Although these phenomena are individually well documented, how their interactions change across operating regimes and influence wake development and power extraction remains insufficiently understood~\cite{Wisner_Yu_JRSE_2024}. Moreover, the computational cost of studying these phenomena with high-fidelity blade-resolved simulations remains prohibitively high~\cite{li20132,LiuEtAl_RE_2019}.

The blade-scale physics originates from the cyclic variation of the relative velocity over each revolution. At relatively low tip-speed ratios ($\lambda \leq 2$), the blade exceeds its static stall angle and undergoes a sequence of shear-layer growth, vortex formation, massive flow separation, and subsequent reattachment~\cite{simao2008visualization,lefouest2022dilemma,lefouest2023timescales}. These processes are characteristic of dynamic stall, whose general features include delayed separation, transient load overshoot, vortex shedding, and pronounced load hysteresis~\cite{mccroskey1981,leishman2006principles,mulleners2012onset}. Moderate dynamic stall can enhance power extraction during the upwind passage by sustaining lift through the shear-layer-growth and vortex-formation stages~\cite{Wisner_Yu_JRSE_2024}. Under deep dynamic stall, however, this enhancement is temporary. After the leading-edge vortex is shed, the lift decreases sharply and the separated flow becomes detrimental to the overall turbine performance~\cite{LiuEtAl_RE_2019, lefouest2023timescales, Wisner_Yu_JRSE_2024}. As the tip-speed ratio increases, the duration and severity of separation decrease, and the blade response transitions from deep stall through moderate or light stall toward an unstalled regime~\cite{LiuEtAl_RE_2019,lefouest2022dilemma,lefouest2023timescales,Wisner_Yu_JRSE_2024}. These operating-regime changes alter not only the instantaneous blade forces but also when useful aerodynamic work is produced during the revolution.

The regime-dependent blade loading is reflected in the structure of the wake. Combined particle image velocimetry (PIV) and large eddy simulation (LES) measurements have shown that VAWT wakes are asymmetric, with the momentum deficit displaced toward the windward side and stronger turbulence and coherent motion developing on the leeward side because of upwind blade stall~\cite{posa2016wake}. Farther downstream, the wake evolves through three distinct regions: a near wake dominated by periodic blade-vortex shedding, a transitional region governed by shear layer instability, and a far wake characterized by bluff-body-like oscillations~\cite{araya2017transition}. The location of these regions and the rate of wake recovery depend on the rotor operating condition and dynamic solidity~\cite{araya2017transition}. These studies explain how blade-scale unsteadiness shapes wake development, but the relationship between instantaneous vortex activity, the mean momentum deficit, and useful power extraction across operating regimes remains less clear.

Blade-resolved CFD provides direct access to the cyclic blade loading and vortex evolution underlying these wake processes. Sliding-mesh unsteady Reynolds-averaged Navier--Stokes (URANS) simulations are widely used because they can reproduce the periodic variation of blade forces at moderate computational cost, although their accuracy depends strongly on the turbulence treatment~\cite{simao2010simulating,lanzafame20142d}. In separated and post-stall flow, 2.5D LES produces more realistic three-dimensional vortex diffusion and more accurate aerodynamic coefficients than two-dimensional URANS~\cite{li20132}. Detailed stereoscopic PIV measurements have consequently become important benchmarks for evaluating the predicted velocity and vorticity fields within the rotor and near wake~\cite{tescione2014near}. Nevertheless, resolving the blade geometry and separated flow over many revolutions remains computationally demanding, particularly when several operating regimes must be examined. This limitation motivates reduced-order representations that retain the essential unsteady loading and wake physics without resolving the blade boundary layer directly.

Actuator-based representations reduce this cost by replacing the explicit blade geometry with distributed body forces. Unlike an actuator disk, the actuator line model (ALM) follows the instantaneous blade positions and can therefore represent azimuthally varying loads and their contribution to the wake~\cite{sorensen2002numerical,bachant2018actuator}. VAWT-specific implementations have demonstrated substantial computational savings while retaining important performance and wake features, although their accuracy remains sensitive to dynamic stall modeling, force projection, and velocity sampling~\cite{bachant2018actuator,melani2021tailoring}. In particular, the Gaussian smoothing width controls both the spatial distribution of the blade force and the velocity induced near the actuator line~\cite{martineztossas2017optimal}. Recent VAWT formulations have consequently introduced dedicated sampling procedures, aerodynamic corrections, and kernel-sensitivity analyses to improve phase-dependent loading and wake prediction~\cite{hezaveh2017simulation,melani2021tailoring}. Preserving the vortical structures generated by this forcing also requires a spatial discretization with sufficiently low numerical dissipation. High-order formulations provide this capability, but their actuator-line applications have focused predominantly on horizontal-axis rotors, leaving VAWT operation dominated by dynamic stall comparatively unexplored~\cite{Frere2016DG_WT,ding2023high,marino2024modelling}.

Against this background, the present study examines how dynamic stall redistributes blade loading between the upwind and downwind passages of a VAWT and how the resulting instantaneous wake relates to mean momentum and power extraction across deep-stall at low tip-speed ratios, near-optimal, and post-optimal (i.e., at high tip-speed ratios) operating regimes. To enable this analysis over a range of tip-speed ratios efficiently, the previously developed high-order source-term formulation~\cite{imran2025high,imran2025development,imran2026} is extended from steady actuator-disk representations to a time-accurate rotating FR/CPR-ALM framework on fixed Cartesian grids. Herein, FR/CPR stands for the flux reconstruction or correction procedure via reconstruction method for high-order CFD simulations~\cite{huynh2007flux,wang2009unifying,ywl14}. The framework enables an unsteady distributed actuator forces representation at subelement resolution, and combines Gaussian force projection, a modified Boeing--Vertol dynamic stall model, and an induction closure based on the double multiple streamtube (DMST) concept. A mesh-resolution criterion is introduced to determine whether the Gaussian width is controlled by the blade chord or the computational mesh. The framework is validated against experimental near-wake measurements and previously reported LES-ALM results, and then applied to compare phase-dependent blade loading across the selected operating regimes and to relate instantaneous wake organization, mean velocity deficit, and power extraction for the deep-stall and near-optimal conditions.

The remainder of the paper is organized as follows. Section~\ref{sec:methodology} presents the FR/CPR-ALM formulation, including actuator force calculation, Gaussian projection, dynamic stall modeling, and induction treatment. Section~\ref{sec:results} first validates the framework and then examines turbine performance, phase-dependent blade loading, and instantaneous and time-averaged wake behavior. Section~\ref{sec:conclusion} summarizes the principal physical and methodological findings.

\section{Methodology}
\label{sec:methodology}
\subsection{FR/CPR discretization}
\label{sec:frcpr}
High-order methods provide low numerical dispersion and dissipation, making them well suited to vortex-dominated flows in which coherent structures must be preserved over long propagation distances~\cite{wang2013high,witherden2016high}. The FR/CPR formulation represents the solution within each element by high-degree polynomials and restores inter-element conservation through correction functions~\cite{huynh2007flux,wang2009unifying,vincent2011new}. With an appropriate choice of correction functions, the formulation recovers the nodal discontinuous Galerkin (DG) method while retaining a compact and computationally efficient implementation. These properties motivate its use here to resolve the unsteady vortical structures generated by the rotating actuator forces.

The flow governed by the two-dimensional compressible Navier--Stokes equations can be written in conservative form as
\begin{equation}
\frac{\partial \mathbf{Q}}{\partial t}
+
\frac{\partial \mathbf{F}}{\partial x}
+
\frac{\partial \mathbf{G}}{\partial y}
=
\mathbf{S},
\label{eq:NS_cons}
\end{equation}
where $\mathbf{Q}$ is the vector of conserved variables, $\mathbf{F} = \mathbf{F}^{\mathrm{inv}}-\mathbf{F}^{\mathrm{vis}}$ and $\mathbf{G} =\mathbf{G}^{\mathrm{inv}}-\mathbf{G}^{\mathrm{vis}}$ are flux vectors in the $x$- and $y$-direction with the superscript ``$\mathrm{inv}$" representing inviscid fluxes, and ``$\mathrm{vis}$" representing viscous fluxes, and $\mathbf{S}$ is the source vector.
The vector of conserved variables is defined as
$\mathbf{Q} = \left(\rho, \rho u, \rho v, E \right)^{\top}$.
Herein, $\rho$ is the density, $(u,v)$ are the Cartesian velocity components, and $E$ is the total energy per unit volume. Closure for calorically a perfect gas is provided by
\begin{equation}
E = \frac{p}{\gamma - 1} + \frac{1}{2}\rho (u^2+v^2),
\end{equation}
where $\gamma$ is the ratio of specific heats.
The definitions of inviscid and viscous fluxes follow our previous works~\cite{imran2025high,imran2025development,imran2026}.

\begin{figure}
\centering
\includegraphics[width=0.75\linewidth]{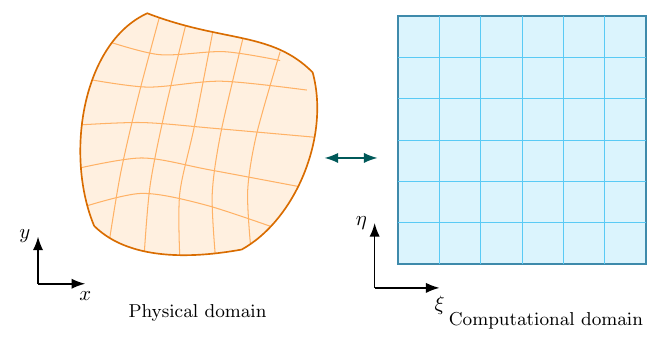}
\caption{Schematic of the coordinate transformation between the physical domain $(x,y)$ and the computational domain $(\xi,\eta)$.}
\label{fig:transformation}
\end{figure}

The FR/CPR discretization is built on a reference element rather than on the physical mesh directly. The physical domain $\Omega$ is first covered by a set of non-overlapping quadrilateral elements $\Omega_e$, i.e.,
$\Omega = \bigcup_{e=1}^{N_e} \Omega_e$, where $N_e$ is the number of elements.
Every element $\Omega_e$ is then mapped onto the standard square
$(\xi,\eta)\in[-1,1]^2$, as illustrated in Fig.~\ref{fig:transformation}. As a result, all operations are carried out on a single fixed reference geometry. Under this mapping the governing system keeps its conservative structure and becomes
\begin{equation}
\frac{\partial \tilde{\mathbf{Q}}}{\partial t}
+
\frac{\partial \tilde{\mathbf{F}}}{\partial \xi}
+
\frac{\partial \tilde{\mathbf{G}}}{\partial \eta}
=
\tilde{\mathbf{S}},
\label{eq:NS_transformed}
\end{equation}
where the transformed state is scaled by the mapping Jacobian,
\begin{equation}
\tilde{\mathbf{Q}} = |J|\mathbf{Q}, \quad 
\tilde{\mathbf{S}} = |J|\mathbf{S},
\end{equation}
and $|J|$ is the Jacobian determinant. The transformed fluxes gather the physical fluxes weighted by the metric terms,
\begin{align}
\tilde{\mathbf{F}} &= |J|\left( \mathbf{F}\,\xi_x + \mathbf{G}\,\xi_y\right), \label{eq:F_flux} \\
\tilde{\mathbf{G}} &= |J|\left( \mathbf{F}\,\eta_x + \mathbf{G}\,\eta_y\right). \label{eq:G_flux}
\end{align}

The calculation of the metric terms in Eqs.~\eqref{eq:F_flux} and \eqref{eq:G_flux} can be found in our previous works~\cite{imran2025high,imran2025development,imran2026}. For completeness, solution and flux approximations in FR/CPR are briefly explained below.

\begin{figure}[h]
\centering
\includegraphics[width=0.75\linewidth]{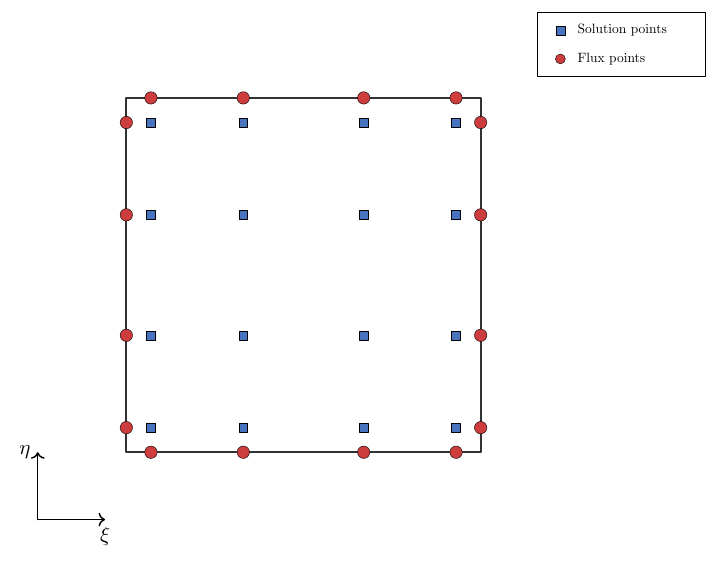}
\caption{Distribution of solution points (blue squares) and flux points (red circles) within a standard quadrilateral reference element for a fourth-order scheme.}
\label{fig:solutionflux}
\end{figure}

Within each reference element the solution is represented by a tensor-product polynomial of degree $N$ in $\xi$ and $\eta$. Using a nodal Lagrange basis, the transformed state is expanded as
\begin{equation}
\tilde{\mathbf{Q}}^h(\xi,\eta,\tau)
=
\sum_{i=1}^{N_s}
\tilde{\mathbf{Q}}_i(\tau)\,
\ell_i(\xi,\eta),
\end{equation}
where the $\ell_i$ are Lagrange polynomials tied to $N_s=(N+1)^2$ solution points. Gauss--Legendre points are used for these nodes, since they combine accurate quadrature with numerical stability. Their layout, together with the flux points introduced below, is shown in Fig.~\ref{fig:solutionflux}.

The fluxes are first sampled at the solution points, which gives an element-local flux polynomial in each direction,
\begin{align}
\tilde{\mathbf{F}}_{\mathrm{loc}}^h(\xi,\eta)
&=
\sum_{i=1}^{N_s}
\tilde{\mathbf{F}}(\tilde{\mathbf{Q}}_i)\,
\ell_i(\xi,\eta), \label{eq:Floc} \\
\tilde{\mathbf{G}}_{\mathrm{loc}}^h(\xi,\eta)
&=
\sum_{i=1}^{N_s}
\tilde{\mathbf{G}}(\tilde{\mathbf{Q}}_i)\,
\ell_i(\xi,\eta), \label{eq:Gloc}
\end{align}
with $\tilde{\mathbf{F}}$ and $\tilde{\mathbf{G}}$ defined in
Eqs.~\eqref{eq:F_flux} and \eqref{eq:G_flux}. Because each element is treated independently, these local polynomials are generally discontinuous at element interfaces and cannot enforce conservation on their own.

Conservation and interelement coupling are restored through a common interface flux. Left and right states are reconstructed at a set of flux points on the element boundaries, and a single-valued flux is then formed from the two sides. For the inviscid part, a Roe-type approximate Riemann solver provides the common flux~\cite{roe1981approximate}, while the common viscous flux is evaluated with the second approach of Bassi and Rebay (BR2)~\cite{Bassi_EtAl_CF_2005}. The gap between the local flux and this common flux at the boundaries is redistributed into the element interior by correction functions. As a result, the corrected polynomials read
\begin{align}
\label{eq:corr1}
\tilde{\mathbf{F}}^h(\xi,\eta)
&=
\tilde{\mathbf{F}}_{\mathrm{loc}}^h(\xi,\eta)
+
\tilde{\mathbf{F}}_{\mathrm{cor}}^h(\xi,\eta), \\
\label{eq:corr2}
\tilde{\mathbf{G}}^h(\xi,\eta)
&=
\tilde{\mathbf{G}}_{\mathrm{loc}}^h(\xi,\eta)
+
\tilde{\mathbf{G}}_{\mathrm{cor}}^h(\xi,\eta).
\end{align}
The corrections $\tilde{\mathbf{F}}_{\mathrm{cor}}^h$ and
$\tilde{\mathbf{G}}_{\mathrm{cor}}^h$ are chosen so that the corrected flux polynomials reproduce the common fluxes exactly at the element boundaries. Radau polynomials are adopted for this purpose, a choice that recovers the DG scheme within the FR/CPR family~\cite{huynh2007flux}.

Differentiating the corrected fluxes and inserting them into
Eq.~\eqref{eq:NS_transformed} produces the semi-discrete update at each solution point,
\begin{equation}
\frac{d\tilde{\mathbf{Q}}_i}{dt}
=
-\left(
\frac{\partial \tilde{\mathbf{F}}^h}{\partial \xi}
+
\frac{\partial \tilde{\mathbf{G}}^h}{\partial \eta}
\right)_i 
+ \tilde{\mathbf{S}}_i.
\label{eq:semi_discrete}
\end{equation}

\bigskip

This can be rewritten in a general form as
\begin{equation}
\label{eq:temporaldis}
\frac{d\mathbf{Q}}{dt}=R(\mathbf{Q},\nabla \mathbf{Q}),
\end{equation}
where $R$ is the spatial residual. Equation~\eqref{eq:temporaldis} is advanced in time with the explicit strong-stability-preserving Runge--Kutta (SSP-RK) scheme~\cite{shu1988total}, which retains the stability of the spatial operator and is well suited to hyperbolic systems. The three-stage form used here advances the solution as
\begin{equation}
\label{eq:SSP-RK}
\begin{cases}
\mathbf{Q}^{(1)}=\mathbf{Q}^{n}+\Delta t \mathbf{R}^n\\[2pt]
\mathbf{Q}^{(2)}=\frac{3}{4}\mathbf{Q}^{n}+\frac{1}{4}\mathbf{Q}^{(1)}+\frac{1}{4}\Delta t \mathbf{R}^{(1)}\\[2pt]
\mathbf{Q}^{n+1}=\frac{1}{3}\mathbf{Q}^{n}+\frac{2}{3}\mathbf{Q}^{(2)}+\frac{2}{3}\Delta t \mathbf{R}^{(2)}
\end{cases}.
\end{equation}

Further details of the solver can be found in our earlier works~\cite{ywl14,wy20}.

\subsection{Actuator line model for vertical-axis wind turbines}
\label{sec:alm}
The ALM method extends the classical theory of blade elements by representing turbine blades as rotating lifting lines embedded within the flow field~\cite{sorensen2002numerical}. 
 
In a two-dimensional formulation, each blade is modeled using a single actuator point, representing a spanwise-uniform blade element. This differs from conventional three-dimensional ALM approaches, where multiple elements are distributed along the blade span. The resulting forces are introduced into the momentum equations as volumetric source terms and are computed from local blade-element aerodynamics. The resulting system can be written as
\begin{equation}
\frac{\partial \mathbf{Q}}{\partial t} + \nabla \cdot \mathbf{F}(\mathbf{Q}) = \mathbf{S}(\mathbf{x}),
\label{eq:NS_with_source}
\end{equation}
where $\mathbf{S}$ represents the actuator-line source vector written as
\begin{equation}
\mathbf{S}(\mathbf{x}) =
\begin{bmatrix} 0 \\ S_x \\ S_y \\ S_e \end{bmatrix}.
\label{eq:source_ALM}
\end{equation}
Herein, $S_x$ and $S_y$ are the body forces acting on fluid in the $x$- and $y$-direction, respectively, and $S_e$ is the resulting work on fluids.

\begin{figure}[h]
\centering
\includegraphics[width=0.90\textwidth]{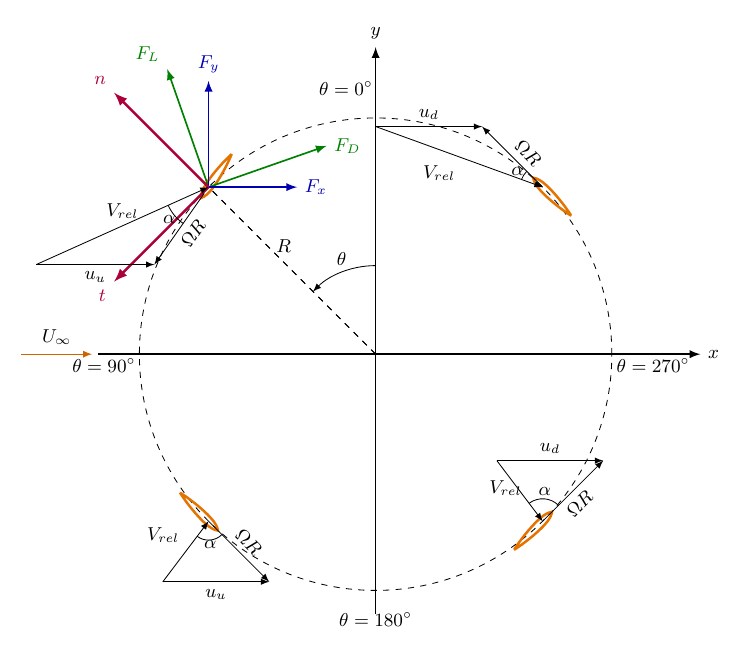}
\caption{Velocity triangles and force decomposition for a rotating blade, drawn
at one azimuthal position per quadrant to illustrate the variation around the
revolution. Force components are shown on the upper-left instance. Blade
positions are schematic.}
\label{fig:vawt_kinematics}
\end{figure}

Consider a horizontal cross-section of a blade rotating with angular velocity $\Omega$ at a fixed radius $R$ from the rotor center, shown in Fig.~\ref{fig:vawt_kinematics}. The  incoming flow is a uniform freestream $\mathbf{U}_\infty$, and the azimuthal angle $\theta$ is measured counter-clockwise from the $+y$ axis. Because the blade extracts momentum from the flow, the velocity it experiences is the reduced local inflow, denoted $u_u$ in the upwind half and $u_d$ in the downwind half following the DMST-based induction closure described in  Sec.~\ref{sec:dmst_formulation}. At each actuator point, a local frame is defined by the radially outward normal $\hat{\mathbf{n}}$ and the tangent $\hat{\mathbf{t}}$, which points in the direction of blade motion. The section travels at velocity $\Omega R\,\hat{\mathbf{t}}$, so the relative velocity experienced by the blade is
\begin{equation}
\mathbf{V}_{\mathrm{rel}} = \mathbf{u}_{\mathrm{loc}} - \Omega R\,\hat{\mathbf{t}},
\label{eq:vrel}
\end{equation}
where $\mathbf{u}_{\mathrm{loc}}$ is the local inflow at the actuator point, equal to $u_u$ or $u_d$ depending on the azimuthal half. Projecting onto the local frame gives the chord-normal and chord-tangential components $V_n = \mathbf{V}_{\mathrm{rel}}\cdot\hat{\mathbf{n}}$ and $V_t = \mathbf{V}_{\mathrm{rel}}\cdot\hat{\mathbf{t}}$, from which the angle of attack is
\begin{equation}
\alpha = \tan^{-1}\!\left(\frac{V_n}{V_t}\right).
\label{eq:aoa}
\end{equation}

The four-quadrant form, together with the chord convention $-\hat{\mathbf{t}}$, resolves $\alpha$ over the full range of incidence swept in one
revolution.

For a vertical-axis turbine, the incidence changes substantially and rapidly over one revolution, and at low to moderate tip-speed ratios, the blade passes well beyond static stall. This drives dynamic stall~\cite{mccroskey1981}, marked by delayed
separation and hysteresis in the aerodynamic loads. A modified Boeing--Vertol model~\cite{gormont1973mathematical} corrects the quasi-steady coefficients. The rate of change of incidence is evaluated once per physical time step by a backward difference,
\begin{equation}
\dot{\alpha} = \frac{\alpha^{n}-\alpha^{n-1}}{\Delta t}.
\end{equation}
The corrected incidences for lift and drag are
\begin{align}
\alpha_L^\ast &= \alpha - \gamma_L\,\kappa
\left(\left|\frac{c\dot{\alpha}}{2\,|\mathbf{V}_{\mathrm{rel}}|}\right|\right)^{1/2}
\frac{\dot{\alpha}}{|\dot{\alpha}|}, \\
\alpha_D^\ast &= \alpha - \gamma_D\,\kappa
\left(\left|\frac{c\dot{\alpha}}{2\,|\mathbf{V}_{\mathrm{rel}}|}\right|\right)^{1/2}
\frac{\dot{\alpha}}{|\dot{\alpha}|},
\end{align}
with empirical coefficients
\begin{equation}
\gamma_L = 1.4 - 6\!\left(0.06-\frac{t_b}{c}\right), \qquad
\gamma_D = 1 - 2.5\!\left(0.06-\frac{t_b}{c}\right), \qquad
\kappa =
\begin{cases}
1.0, & \dot{\alpha}\ge 0,\\
0.5, & \dot{\alpha}<0,
\end{cases}
\end{equation}
where $t_b/c$ is the thickness ratio, equal to $0.18$ for the NACA~0018 section
used here, giving $\gamma_L = 2.12$ and $\gamma_D = 1.30$. For the symmetric
airfoil, the zero-lift angle is $\alpha_0 = 0$. The corrected coefficients are
\begin{equation}
C_L^\ast = \left(\frac{\alpha}{\alpha_L^\ast-\alpha_0}\right)C_L(\alpha_L^\ast),
\qquad
C_D^\ast = C_D(\alpha_D^\ast),
\end{equation}
with $C_L(\cdot)$ and $C_D(\cdot)$ from tabulated static polars. When
$|\alpha_L^\ast-\alpha_0|$ falls below a small threshold, the ratio is bypassed and
$C_L^\ast = C_L(\alpha_L^\ast)$ is used, which removes the singularity as
$\alpha_L^\ast\to\alpha_0$. This construction preserves the static polar under
quasi-steady conditions and introduces rate-dependent hysteresis through
$\dot{\alpha}$ during the rapid incidence changes typical of VAWT operation.

The sectional lift and drag magnitudes are
\begin{equation}
L = \tfrac{1}{2}\rho_\infty |\mathbf{V}_{\mathrm{rel}}|^2\,c\,C_L^\ast, \qquad
D = \tfrac{1}{2}\rho_\infty |\mathbf{V}_{\mathrm{rel}}|^2\,c\,C_D^\ast,
\end{equation}
and act along the wind-aligned unit vectors
$\hat{\mathbf{e}}_D = \mathbf{V}_{\mathrm{rel}}/|\mathbf{V}_{\mathrm{rel}}|$ and
$\hat{\mathbf{e}}_L = \hat{\mathbf{e}}_z\times\hat{\mathbf{e}}_D$. Therefore, the aerodynamic force on the blade, per unit span, is
\begin{equation}
\label{eq:Fb}
\mathbf{F}_b = L\,\hat{\mathbf{e}}_L + D\,\hat{\mathbf{e}}_D.
\end{equation}
For the evaluation of torque and diagnostic quantities, the aerodynamic force is decomposed into tangential and normal components as
\begin{equation}
F_t = L\sin\alpha - D\cos\alpha, \qquad
F_n = L\cos\alpha + D\sin\alpha,
\end{equation}
with $F_t$ positive in the power-generating direction. The per-blade contribution
to rotor torque is $Q_b = R\,F_t$.

The discrete blade forces are returned to the flow as smooth volumetric source terms over a finite Gaussian support. When adequately resolved, the support distributes each force across multiple solution points rather than concentrating it within the element containing the actuator point. The present implementation uses an isotropic Gaussian kernel centered on
each blade. For a point $\mathbf{x}$ with displacement
$\Delta\mathbf{x} = \mathbf{x}-\mathbf{x}_b$ from blade $b$, the kernel is
\begin{equation}
G_b(\mathbf{x}) = \frac{1}{\pi\,\varepsilon^{2}}
\exp\!\left(-\frac{|\Delta\mathbf{x}|^{2}}{\varepsilon^{2}}\right),
\label{eq:gauss_iso}
\end{equation}
which integrates to unity over $\mathbb{R}^2$. The width $\varepsilon$ is set by
the larger value of a mesh measure and a chord measure,
\begin{equation}
\varepsilon = \max\!\left(\frac{2h}{p},\ \frac{c}{\kappa_c}\right),
\label{eq:eps_choice}
\end{equation}
where $h$ is the local element size, $p$ is the solution polynomial degree, and $\kappa_c$ is a chord-fraction constant, set to $\kappa_c = 4.3$ in all simulations reported here. The mesh term, $2h/p$ keeps the kernel resolvable on the grid, while the chord floor, $c/\kappa_c$ prevents the support from shrinking below the blade scale. For efficiency the kernel support is truncated at $r_c = 4\varepsilon$.

By Newton's third law the force on the fluid is equal and opposite to the
aerodynamic force on each blade. Summing the kernel-weighted reaction of every
blade gives the source term vector,
\begin{equation}
\mathbf{S}(\mathbf{x}) =
\begin{bmatrix} 0 \\ S_x \\ S_y \\ S_e \end{bmatrix}
= \sum_{b=1}^{N_B} G_b(\mathbf{x})
\begin{bmatrix}
0 \\
-F_{x,b} \\
-F_{y,b} \\
-F_{e,b}
\end{bmatrix},
\label{eq:source_all}
\end{equation}
where $F_{x,b}$ and $F_{y,b}$ are the components of the aerodynamic force $\mathbf{F}_b$ on blade $b$ from Eq.~\eqref{eq:Fb}, and $-F_{e,b} = -\mathbf{F}_b\cdot\mathbf{u}$ is the corresponding rate of work done on the fluid. The momentum and energy forcing share the same projection $G_b$, and are therefore consistent by construction.

The instantaneous aerodynamic torque about the rotation axis (per unit span) and the mechanical power are computed as
\begin{equation}
\mathcal{T} = \sum_{b=1}^{N_B}\left(\mathbf{r}_b \times \mathbf{F}_b\right)
\cdot\hat{\mathbf{e}}_z,
\qquad
\mathcal{P}=\Omega\,\mathcal{T},
\label{eq:torque_power}
\end{equation}
where $\mathbf{r}_b$ is the blade position vector in the rotor plane. For the two-dimensional simulations reported herein, the reference swept area per unit span is $A_{\mathrm{ref}}=D=2R$, corresponding to
the projected frontal area per unit span. The power coefficient is then expressed as

\begin{equation}
C_P = \frac{\mathcal{P}}{\tfrac12\,\rho_\infty U_\infty^3 A_{\mathrm{ref}}}.
\label{eq:cp}
\end{equation}
The source terms in Eq.~\eqref{eq:source_all} are evaluated at the element quadrature points and assembled into the semi-discrete residual as volumetric forcing terms. The isotropic smoothing in Eq.~\eqref{eq:gauss_iso}, with the grid-dependent width in Eq.~\eqref{eq:eps_choice}, supports stable time integration and provides a resolved representation of the rotor-induced momentum deficit and wake development on the meshes considered. The present two-dimensional formulation provides a controlled and computationally efficient setting for isolating sectional blade-loading mechanisms and comparing their wake response across multiple operating conditions. Each blade is represented by a single actuator point in the computational plane, and its aerodynamic force is distributed through the Gaussian kernel. The blade-scale aerodynamic response is supplied by the tabulated coefficients and the modified Boeing--Vertol dynamic stall model, while the FR/CPR solution resolves the resulting unsteady flow and wake development. This formulation therefore enables systematic examination of the relationship among phase-dependent loading, instantaneous wake organization, mean velocity deficit, and power extraction over a range of tip-speed ratios. The model does not include finite-span effects, blade-tip vortices, spanwise transport, vortex stretching, or three-dimensional vortex breakdown. These mechanisms can influence the detailed evolution and recovery of a finite-span turbine wake. Accordingly, the wake results are interpreted as comparative trends among the two-dimensional operating conditions examined. Their applicability to finite-span turbines is assessed through the available experimental and three-dimensional LES-ALM comparisons, while a direct three-dimensional extension is reserved for future work.

\section{Results and Discussion}
\label{sec:results}
\subsection{Numerical setup}
\label{sec:computational_setup}
We perform all simulations using a high-order FR/CPR solver coupled with the ALM on fixed, structured Cartesian meshes. The computational domain is a two-dimensional rectangular region defined by $x,y \in [-20D,\,20D]$, where $D=2R$ denotes the rotor diameter. The turbine is located at the domain center $(0,0)$ and is represented solely through volumetric ALM forcing. A uniform freestream velocity $U_\infty$ is prescribed at the inlet, a zero-gradient condition is applied at the outlet, and slip conditions are imposed on the top and bottom boundaries. The domain size is chosen so that the near-wake region is not influenced by the
outer boundaries.

\begin{figure}[!htbp]
    \centering
    \begin{minipage}{0.48\linewidth}
        \centering
        \includegraphics[width=\linewidth]{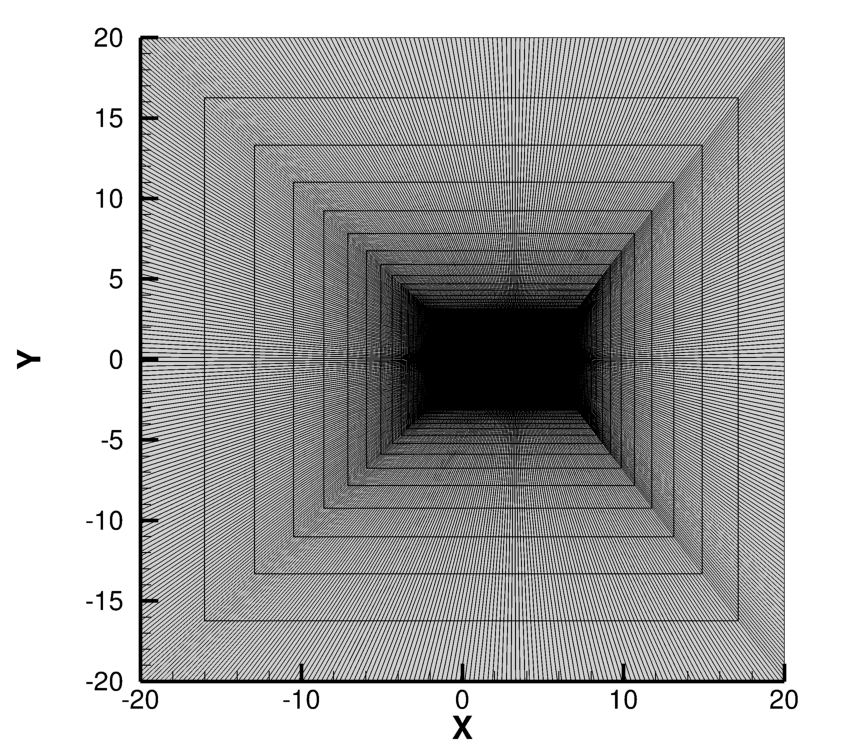}
        \par\smallskip
        (a) Full computational domain
        \label{fig:mesh_full}
    \end{minipage}
    \hfill
    \begin{minipage}{0.48\linewidth}
        \centering
        \includegraphics[width=\linewidth]{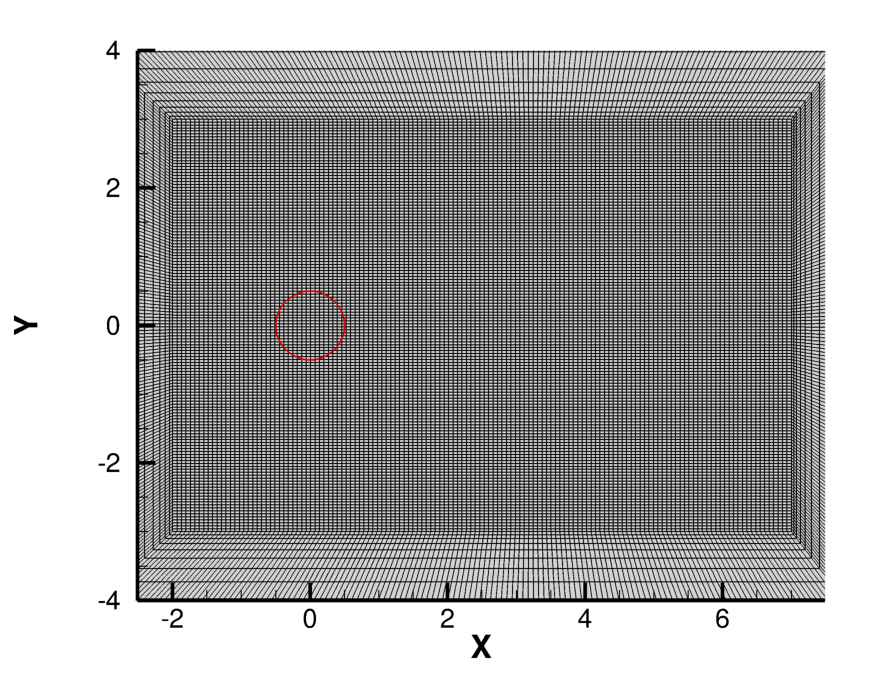}
        \par\smallskip
        (b) Zoomed view of the turbine-center region
        \label{fig:mesh_zoom}
    \end{minipage}

    \caption{Computational mesh with local refinement in the turbine region. The red circle of radius $0.5\,\mathrm{m}$ indicates the rotor location
    and is shown for visualization only; no physical geometry is present in the mesh.}
    \label{fig:mesh_overview}
\end{figure}

To accurately capture the actuator forcing and near-wake dynamics, we introduce a locally refined rectangular region around the turbine spanning $x \in [-2D,\,7D]$ and $y \in [-3D,\,3D]$. Outside this
region, the mesh is gradually coarsened toward the far-field boundaries to reduce computational cost. Figure~\ref{fig:mesh_overview} shows the overall mesh and a zoomed view of the refined turbine region; the rotor location is indicated by a circle for reference only, as no physical blade geometry is present.

The solution is advanced from a uniform freestream initial condition with a constant time step $\Delta t=10^{-4}\,\mathrm{s}$, and time-averaged statistics are collected after a statistically periodic state is reached, typically within $10$-$15$ rotor revolutions. Blade aerodynamic coefficients are obtained from tabulated static data~\cite{sheldahl1981aerodynamic} and corrected for unsteady effects using the Boeing--Vertol dynamic stall model (Sec.~\ref{sec:alm}).
The effective blade inflow is supplied by the DMST-based induction closure described in Sec.~\ref{sec:cp_lambda}.

\subsection{Model validation}
\label{sec:validation}
The present FR/CPR-ALM framework is validated against the experimental near-wake measurements of Bachant and
Wosnik~\cite{bachant2013performance} and the ALM-LES results of Hezaveh et al.~\cite{hezaveh2017simulation}. The reference configuration is a three-bladed vertical-axis turbine of diameter
$D=1\,\mathrm{m}$ with chord $c=0.14\,\mathrm{m}$, operating at a diameter-based Reynolds number, $Re_D = U_\infty D/\nu \approx 10^{6}$, matching the reference experiments. The solidity is defined here as $\sigma = N_b c/(2\pi R) = 3\times0.14/(2\pi\times0.5)\approx0.13$. The validation is carried out at $\lambda=1.9$, the tip-speed ratio at which Bachant and Wosnik~\cite{bachant2013performance} report near-wake measurements.

\begin{figure}[h]
\centering
\includegraphics[width=0.75\textwidth]{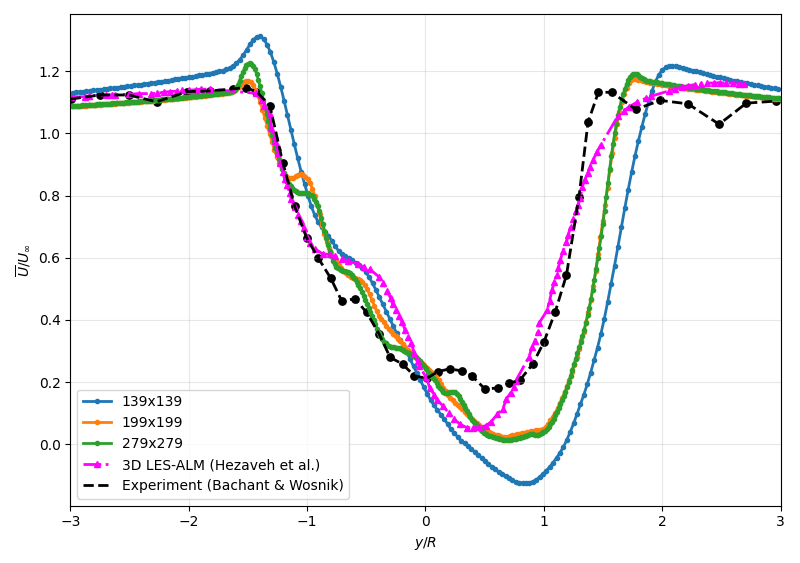}
\caption{Normalized mean streamwise velocity profile $\overline{U}/U_\infty$ as a function of $y/R$ at $x/D=1$ for three grid resolutions, compared with the
experimental measurements of Bachant and Wosnik~\cite{bachant2013performance} and the ALM-LES results of Hezaveh et al.~\cite{hezaveh2017simulation}.}
\label{fig:wake_u}
\end{figure}

For the validation configuration, the inner refined region is discretized using $139\times139$, $199\times199$, and $279\times279$ elements, corresponding to approximately 29k, 62k, and 129k total cells, respectively. Figure~\ref{fig:wake_u} compares the normalized mean streamwise velocity $\overline{U}/U_\infty$ at $x/D=1$ for the three mesh levels against the experimental and ALM-LES reference data. All three numerical solutions reproduce the overall magnitude, lateral extent, and transverse asymmetry of the near-wake velocity deficit. The asymmetry arises from the unequal phase-dependent loading and blade--wake interaction during the upwind and downwind passages, which redistribute momentum across the two sides of the rotor. Its presence in both the experimental and three-dimensional ALM-LES profiles further identifies it as a rotor-induced flow feature. The more noticeable differences among the numerical profiles occur over $y/R\in[0,1]$, demonstrating that localized wake quantities retain
sensitivity to spatial resolution. The detailed magnitude and transverse position of the deficit are influenced by the Gaussian projection, spatial resolution, induction treatment, and two-dimensional representation.

In the adopted coupling, the effective upwind and downwind blade velocities used in the aerodynamic-force calculation are supplied by the DMST-based induction closure. The blade-loading histories are therefore evaluated consistently across the three CFD grids, while grid refinement changes the spatial representation and downstream evolution of the projected forcing in the resolved wake. The role of the DMST-based induction closure and its relationship to the CFD-resolved velocity are described in Sec.~\ref{sec:induction_motivation}.

Based on this three-grid assessment, the coarse grid is used as the common production mesh for the operating-condition sweep. Consequently, the wake comparisons emphasize the broad-scale deficit and wake organization reproduced across the three grid levels; localized velocity and vorticity structures are interpreted qualitatively.

\subsection{Gaussian kernel resolution criterion}
\label{sec:eps_criterion}
The isotropic Gaussian width in Eq.~\eqref{eq:eps_choice} is defined as
$\varepsilon=\max(2h/p,\,c/\kappa_c)$. As the mesh is refined, the
mesh-based value $2h/p$ decreases. Once it becomes smaller than the
chord-based value $c/\kappa_c$, the Gaussian width is determined by the
blade chord rather than the mesh spacing. This condition,
$c/\kappa_c>2h/p$, gives the resolution criterion
\begin{equation}
  h < \frac{c\,p}{2\,\kappa_c}.
  \label{eq:h_criterion}
\end{equation}
For the present geometry ($c=0.14\,\mathrm{m}$, $p=3$, $\kappa_c=4.3$),
Eq.~\eqref{eq:h_criterion} requires $h<0.049\,\mathrm{m}$. Table~\ref{tab:kernel} lists the kernel width and its controlling factor for each mesh level.

\begin{table}[h]
\caption{Isotropic Gaussian kernel width and controlling factor for each mesh level. $h$ is the inner-region element size, $h/p$ the effective spacing at order $p=3$, and $\varepsilon=\max(2h/p,\,c/\kappa_c)$ the kernel width. ``Chord'' denotes the chord-controlled regime and ``Mesh'' the mesh-controlled regime
($c=0.14\,\mathrm{m}$, $\kappa_c=4.3$, so $c/\kappa_c=0.033\,\mathrm{m}$).}
\label{tab:kernel}
\begin{ruledtabular}
\begin{tabular}{lccccc}
Mesh & Cells & $h$\,(m) & $h/p$\,(m) & $\varepsilon$\,(m) & Regime \\
\hline
Coarse ($139^2$)  & 29k  & 0.065 & 0.022 & 0.043 & Mesh  \\
Medium ($199^2$)  & 62k  & 0.045 & 0.015 & 0.033 & Chord \\
Fine   ($279^2$)  & 129k & 0.032 & 0.011 & 0.033 & Chord \\
\end{tabular}
\end{ruledtabular}
\end{table}

For the validation configuration, the mesh term
$2h/p=0.043\,\mathrm{m}$ governs on the coarse mesh. The kernel is
therefore mesh-controlled on the coarse mesh, whereas its width settles to the chord floor $c/\kappa_c=0.033\,\mathrm{m}$ on the intermediate and fine validation meshes. In either regime, the support radius $r_c=4\varepsilon$ spans approximately one chord or more. These values establish the spatial support over which the actuator force is projected on each mesh. In the present coupling, the DMST model supplies the effective blade inflow used in the aerodynamic-force calculation, while the Gaussian width controls the spatial distribution of the projected force and the resulting resolved wake.

\subsection{Power coefficient and induction modeling}
\label{sec:cp_lambda}
The power coefficient as a function of  tip-speed ratio is the primary performance metric of the turbine. Its accurate prediction requires both physically consistent blade loading supplied by the FR/CPR-ALM with Boeing--Vertol correction and an effective inflow velocity that accounts for rotor induction at the blade.

\subsubsection{Motivation for an explicit induction model}
\label{sec:induction_motivation}
Accurate evaluation of the actuator-line forces requires an effective incident velocity that includes the rotor-scale induction while excluding disturbances associated with the regularized blade force itself. One approach is to obtain this velocity directly from the CFD solution at or near the actuator point. Although sampling at the center of an ideal Gaussian lift distribution has a theoretical basis, practical implementations remain coupled to the interpolation procedure, sampling location, force-projection width, and spatial and temporal resolution~\cite{martineztossas2017optimal,xie2021actuator,zormpa2025effect,muscari2024effective}. These choices can influence the evaluated blade-relative velocity and consequently the predicted rotor power. This sensitivity is also relevant to VAWT operating-condition sweeps. Bachant et al.~\cite{bachant2018actuator} sampled the local velocity at the blade quarter-chord from CFD simulations and obtained reasonable agreement through the optimal operating range, but overpredicted the power coefficient at higher tip-speed ratios. The authors did not attribute this discrepancy solely to velocity sampling, identifying aerodynamic-model and parasitic-loss effects as additional possible sources.

For the present tip-speed-ratio sweep, the double multiple streamtube model is therefore used to provide a consistent specification of the rotor-induced blade inflow, while leaving the sensitivity to local velocity sampling for future investigation. The DMST approach separates the rotor into upwind and downwind passages and accounts for the reduced velocity encountered after the flow passes through the upstream portion of the rotor~\cite{paraschivoiu2002,paraschivoiu1988double}. The resulting velocities are used to evaluate the relative velocity of the blade and aerodynamic forces, while the FR/CPR solution evolves the unsteady flow and wake generated by the projected actuator forces.

\subsubsection{Double multiple streamtube induction model}
\label{sec:dmst_formulation}
The DMST model~\cite{paraschivoiu2002} divides the rotor disk into an upwind half cycle ($\theta\in[0^\circ,180^\circ]$) and a downwind half cycle
($\theta\in(180^\circ,360^\circ]$), each treated as an actuator disk. The effective inflow velocity in each half cycle is
\begin{align}
u_u &= U_\infty(1-a_u), \label{eq:dmst_upwind}\\
u_d &= U_\infty(1-2a_u), \label{eq:dmst_downwind}
\end{align}
where the Rankine-Froude momentum theorem gives the downwind condition as twice the upwind induction at the disk plane. The resulting upwind and downwind velocities define the corresponding
velocity triangles and are used to evaluate the angle of attack, relative velocity, and sectional aerodynamic forces during each blade passage. The calculated blade forces are subsequently projected onto the FR/CPR
solution through the Gaussian kernel, and the solver advances the resulting unsteady flow and wake.

The upwind induction factor $a_u$ is obtained from the rotor thrust coefficient $C_T$ via momentum theory with the Glauert correction for heavily loaded rotors~\cite{Glauert1935}:
\begin{equation}
a_u =
\begin{cases}
\dfrac{1-\sqrt{1-C_T}}{2} & C_T \leq 0.96, \\[8pt]
\dfrac{1+\sqrt{1+12\,C_T}}{6} & C_T > 0.96.
\end{cases}
\label{eq:glauert}
\end{equation}
The factor $a_u$ is updated once per rotor revolution from the time-averaged $C_T$ of the preceding revolution, with under-relaxation to suppress oscillations at high tip-speed ratios:
\begin{equation}
a_u^{(n+1)} = 0.9\,a_u^{(n)} + 0.1\,a_u^\mathrm{target}.
\label{eq:relax}
\end{equation}
Starting from $a_u=0$, the induction factor converges to a stationary value within approximately ten revolutions at all tip-speed ratios considered.

\subsubsection{Power coefficient curve}
\label{sec:cp_results}
\begin{figure}[h]
\centering
\includegraphics[width=0.85\textwidth]{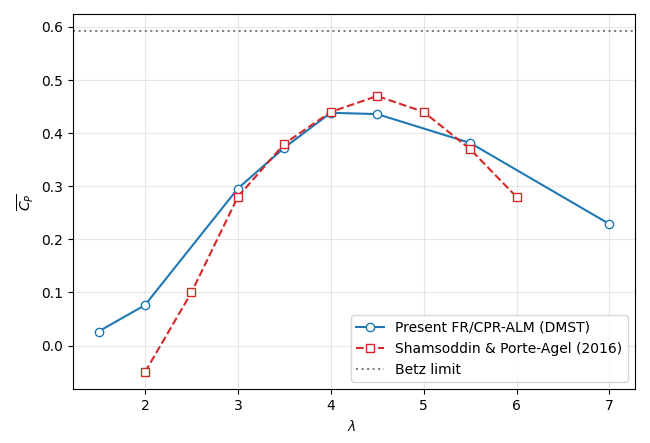}
\caption{Power coefficient $C_P$ as a function of tip-speed ratio, $\lambda=\Omega R/U_\infty$ for the present FR/CPR-ALM with DMST induction model and Boeing--Vertol dynamic stall, compared with the 3D LES-ALM reference of Shamsoddin and Port\'{e}-Agel~\cite{shamsoddin2016large}.}
\label{fig:cp_lambda}
\end{figure}

Figure~\ref{fig:cp_lambda} presents the time-averaged power coefficient as a function of tip-speed ratio obtained with the DMST induction model and the Boeing--Vertol dynamic stall model, together with the 3D LES-ALM results of
Shamsoddin and Port\'{e}-Agel~\cite{shamsoddin2016large} for comparison. Each point is averaged over the final $15\,\mathrm{s}$ of a $30\,\mathrm{s}$ run.

The power coefficient rises from $C_P=0.027$ at $\lambda=1.5$ to a peak of $C_P=0.443$ at $\lambda=4.0$, then declines to $0.229$ at $\lambda=7.0$, which reproduces the characteristic single-peaked $C_P(\lambda)$ behavior of a vertical-axis turbine. The predicted peak power coefficient of $0.443$ is approximately $5.7\%$ lower than the 3D LES-ALM reference peak of $0.47$, and the two curves closely follow each other over the range $\lambda=3.5$--$5.5$ that brackets the optimum. The present peak occurs at $\lambda=4.0$ rather than the reference value of $\lambda=4.5$ predicted by the 3D LES-ALM simulation.

At low tip-speed ratios ($\lambda\leq3.0$), the present results lie above the 3D reference. This agrees with the observation of Liu \textit{et al.}~\cite{LiuEtAl_RE_2019}, who found that the power coefficients predicted at low tip-speed ratios by two-dimensional simulations are much larger than those predicted by three-dimensional simulations. The discrepancy is most pronounced in the stall-dominated regime and diminishes toward the optimum, where the flow remains largely attached, and the two- and three-dimensional predictions come into closer agreement. Table~\ref{tab:cp_results} summarizes the results. The absolute standard deviation $\sigma_{C_P}$ reaches its largest values at $\lambda=3.0$ and 3.5, where substantial cyclic loading persists as the mean power increases, and decreases toward the optimal and post-optimal conditions. Relative to the mean power coefficient, however, the fluctuations remain particularly large at the lowest tip-speed ratios.

\begin{table*}[h]
\caption{Time-averaged power coefficient $\langle C_P\rangle$ and standard deviation $\sigma_{C_P}$ from the present FR/CPR-ALM with DMST induction model and Boeing--Vertol dynamic stall, with the number of rotor revolutions in the averaging window. Results averaged over the final $15\,\mathrm{s}$ of a
$30\,\mathrm{s}$ run. Here $\sigma_{C_P}$ denotes the standard deviation of the instantaneous power coefficient about its time-averaged value over the averaging window.}
\label{tab:cp_results}
\begin{ruledtabular}
\begin{tabular}{lcccc}
$\lambda$ & $\langle C_P\rangle$ & $\sigma_{C_P}$ & Revolutions &
$C_P$ (3D LES-ALM) \\
\hline
1.5 & 0.027 & 0.027 &  7.2 & --   \\
2.0 & 0.075 & 0.078 &  9.5 & $-0.05$ \\
3.0 & 0.296 & 0.137 & 14.3 & 0.28  \\
3.5 & 0.376 & 0.128 & 16.7 & 0.38  \\
\textbf{4.0} & \textbf{0.443} & 0.072 & 19.1 & 0.44 \\
4.5 & 0.439 & 0.073 & 21.5 & 0.47  \\
5.5 & 0.382 & 0.077 & 26.3 & 0.37  \\
7.0 & 0.229 & 0.089 & 33.4 & --   \\
\end{tabular}
\end{ruledtabular}
\end{table*}

\subsection{Effect of the Boeing--Vertol dynamic stall correction}
\label{sec:dynstall}
The influence of the Boeing--Vertol correction is isolated by comparing simulations that use the static airfoil polar directly against those that apply the dynamic stall model, at three representative tip-speed ratios, i.e., $\lambda=1.5$, $3.5$ and $5.5$. These conditions represent deep dynamic stall, light stall near the optimal range, and post-optimal attached operation, respectively. The corresponding angle-of-attack histories are documented in Sec.~\ref{sec:kinematics}.

\subsubsection{Lift hysteresis}
\label{sec:cl_hysteresis}

\begin{figure}[h]
\centering
\includegraphics[width=\textwidth]{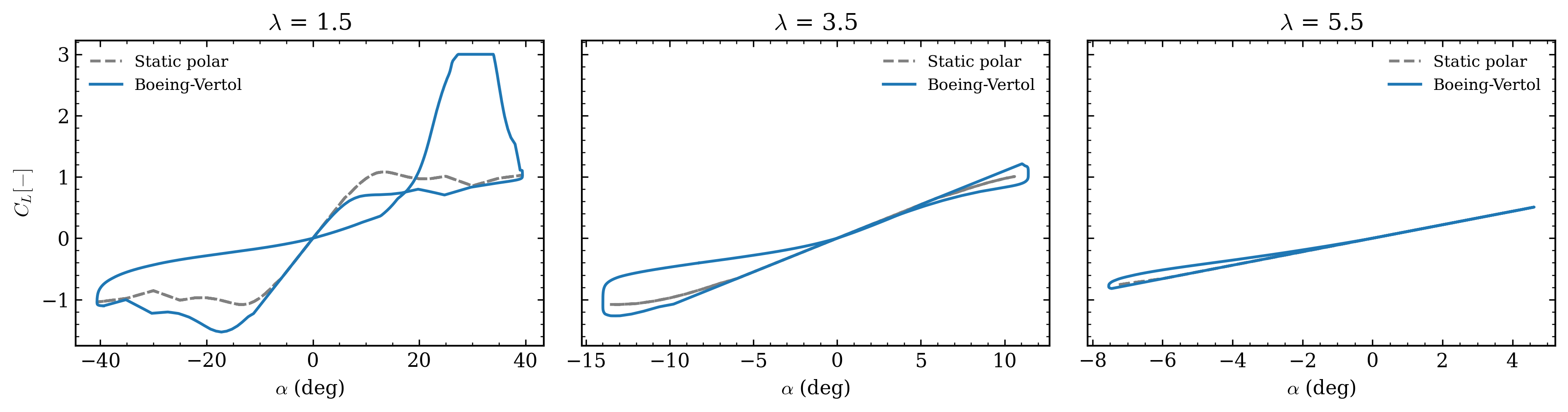}
\caption{Phase-averaged $C_L(\alpha)$ trajectory over one rotor revolution at $\lambda=1.5$ (left), $\lambda=3.5$ (center), and $\lambda=5.5$ (right).}
\label{fig:cl_alpha_ds}
\end{figure}

Figure~\ref{fig:cl_alpha_ds} shows the phase-averaged $C_L(\alpha)$ trajectory over one rotor revolution at the three tip-speed ratios. The static polar is traced reversibly as the angle of attack increases and decreases, whereas the Boeing--Vertol result forms a hysteresis loop characteristic of dynamic stall.

At $\lambda=1.5$, the angle of attack reaches approximately $\pm40^\circ$, far beyond the static stall angle, and the dynamic stall loop is correspondingly large. On the upstroke, the correction represents the delayed-stall effect and sustains lift well above the static maximum, while the modified incidence drives the lift coefficient toward its maximum modeled value near the peak of the excursion. On the downstroke, the lift recovers along a markedly lower branch. The resulting separation between the upstroke and downstroke branches is therefore substantial under this condition. At $\lambda=3.5$, the angle of attack remains within $-14^\circ$ to $+11^\circ$, near the static stall boundary, and the hysteresis loop becomes correspondingly narrower. At $\lambda=5.5$, the incidence is further restricted to approximately $-8^\circ$ to $+5^\circ$. The dynamic and static trajectories then nearly overlap, with only a narrow hysteresis loop. The three conditions demonstrate the progressive weakening of the modeled unsteady lift response as the blade motion changes from deep dynamic stall to light stall and finally to predominantly attached flow.

\subsubsection{Tangential force and torque}
\label{sec:ft_dynstall}
\begin{figure}[h]
\centering
\includegraphics[width=\textwidth]{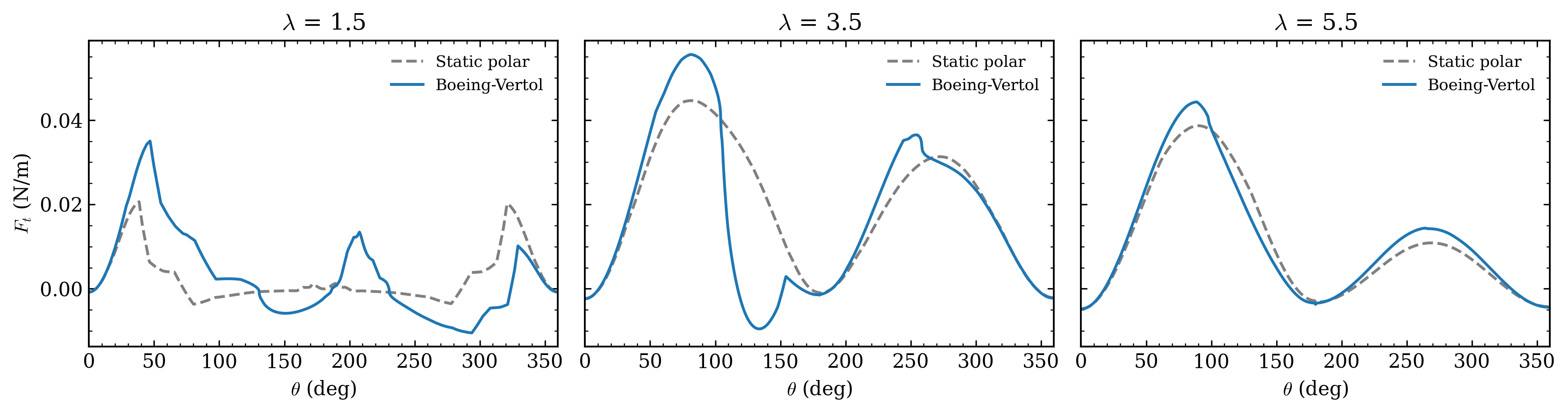}
\caption{Phase-averaged tangential force $F_t(\theta)$ per unit span over one rotor revolution at $\lambda=1.5$ (left), $\lambda=3.5$ (center), and $\lambda=5.5$ (right).}
\label{fig:ft_theta_ds}
\end{figure}

The aerodynamic consequence of the lift hysteresis appears directly in the tangential force $F_t(\theta)$ shown in Fig.~\ref{fig:ft_theta_ds}, which governs the instantaneous torque. At $\lambda=1.5$, the Boeing--Vertol correction raises the tangential force peak in the upwind pass to roughly $0.035\,\mathrm{N/m}$, compared to about $0.020\,\mathrm{N/m}$ for the static polar, and introduces additional structure associated with the modeled delayed-stall response, including a secondary peak in the downwind pass near $\theta\approx200^\circ$. The corrected loading shows regions of negative $F_t$ over part of the cycle, which reflect the deep-stall excursions. At $\lambda=3.5$, the loading is organized into two clean lobes corresponding to the upwind and downwind passes. The Boeing--Vertol peak in the upwind pass ($\approx0.055\,\mathrm{N/m}$) exceeds the static value ($\approx0.045\,\mathrm{N/m}$), consistent with the modest upstroke overshoot seen in the lift hysteresis. At $\lambda=5.5$, the loading retains two smooth lobes, with a stronger upwind contribution and a weaker downwind contribution. The peak tangential force is lower than that at $\lambda=3.5$, consistent with the smaller incidence excursion and the post-optimal reduction in power shown in Fig.~\ref{fig:cp_lambda}.

The net effect of the correction becomes clear when the loading in
Fig.~\ref{fig:ft_theta_ds} is separated into the upwind pass
($\theta\in[0^\circ,180^\circ]$) and the downwind pass
($\theta\in(180^\circ,360^\circ]$) and $F_t$ is averaged over each half. At $\lambda=1.5$  and $\lambda=3.5$, the change in the cycle-averaged loading arises predominantly during the upwind pass, and its effect reverses between the two regimes. At $\lambda=1.5$, the upwind mean tangential force rises from about $0.0025\,\mathrm{N/m}$ for the static polar to $0.0072\,\mathrm{N/m}$ with the correction, an increase of nearly a factor of three, while the downwind mean changes little and turns slightly negative. At $\lambda=3.5$, the upwind mean falls from about $0.0245\,\mathrm{N/m}$ to $0.0189\,\mathrm{N/m}$, a reduction of
roughly $23\%$, and the downwind mean is again barely affected. At $\lambda=5.5$, the upwind mean changes only slightly, while the downwind mean increases from approximately $0.0036\,\mathrm{N/m}$ to $0.0053\,\mathrm{N/m}$.

This upwind localization explains the opposite signs of the two effects. At $\lambda=1.5$, the blade sweeps far beyond the static stall angle during the upwind pass. The delayed separation represented by the Boeing--Vertol correction temporarily sustains lift and recovers tangential force that the static polar does not capture. Following this transient lift enhancement, the tangential loading decreases and becomes negative over part of the cycle. The correction therefore increases the predicted cycle-averaged torque relative to the static-polar result, but the net power extraction remains very low under this deep stall condition. At $\lambda=3.5$, the blade remains closer to the attached-flow regime during the upwind pass, where the static polar overpredicts the quasi-steady lift; the correction removes this excess and lowers the upwind force. The downwind pass contributes little to the change in either regime because the blade operates in the reduced-momentum flow left by the upwind passage. At $\lambda=5.5$, the smaller difference between the two loading histories reflects the weaker influence of the correction under predominantly attached operation. The predominance of the upwind contribution is consistent with previous experimental and numerical observations that the principal dynamic-stall-vortex development and associated transient tangential-force enhancement occur during the upstream blade motion~\cite{lefouest2022dilemma,Wisner_Yu_JRSE_2024}. Consistent with those studies, the present results distinguish this temporary loading enhancement from the overall turbine performance: at $\lambda=1.5$, the transient increase during part of the upwind passage does not compensate for the subsequent loss of favorable loading over the complete revolution.

\subsubsection{Effect on power coefficient}
\label{sec:dynstall_cp}
Table~\ref{tab:dynstall_cp} quantifies the effect of the correction on the time-averaged power coefficient at three tip-speed ratios spanning the deep stall, light stall, and attached regimes. At $\lambda=1.5$, the correction increases $C_P$ by $23\%$ relative to the static-polar result because delayed separation temporarily sustains tangential force during part of the upwind passage. However, the absolute increase is only $0.0050$, from $C_P=0.0217$ to $0.0267$, confirming that deep stall remains detrimental to the overall turbine performance. At $\lambda=3.5$, the correction decreases $C_P$ by $11\%$ by removing the overprediction produced by the static polar in light stall. At $\lambda=5.5$, where the flow is tightly attached, the effect is small ($+7\%$) and reflects only mild unsteady hysteresis. These results demonstrate that the influence of the dynamic stall correction is regime dependent and must be distinguished from the overall aerodynamic benefit or detriment of the corresponding stall regime.

\begin{table}[h]
\caption{Time-averaged power coefficient with and without the Boeing--Vertol dynamic stall correction at three tip-speed ratios. $\Delta C_P$ is the signed difference.}
\label{tab:dynstall_cp}
\begin{ruledtabular}
\begin{tabular}{lcccc}
$\lambda$ & $C_P$ (static) & $C_P$ (Boeing--Vertol) &
$\Delta C_P$ & Change (\%) \\
\hline
1.5 & 0.0217 & 0.0267 & $+0.0050$ & $+23.0$ \\
3.5 & 0.4212 & 0.3761 & $-0.0451$ & $-11.0$ \\
5.5 & 0.3568 & 0.3815 & $+0.0247$ & $+7.0$  \\
\end{tabular}
\end{ruledtabular}
\end{table}

\subsection{Blade kinematics and flow fields}
\label{sec:flowfield}
\subsubsection{Angle of attack and its rate}
\label{sec:kinematics}

\begin{figure}[h]
    \centering
    \begin{minipage}{0.48\linewidth}
        \centering
        \includegraphics[width=\linewidth]{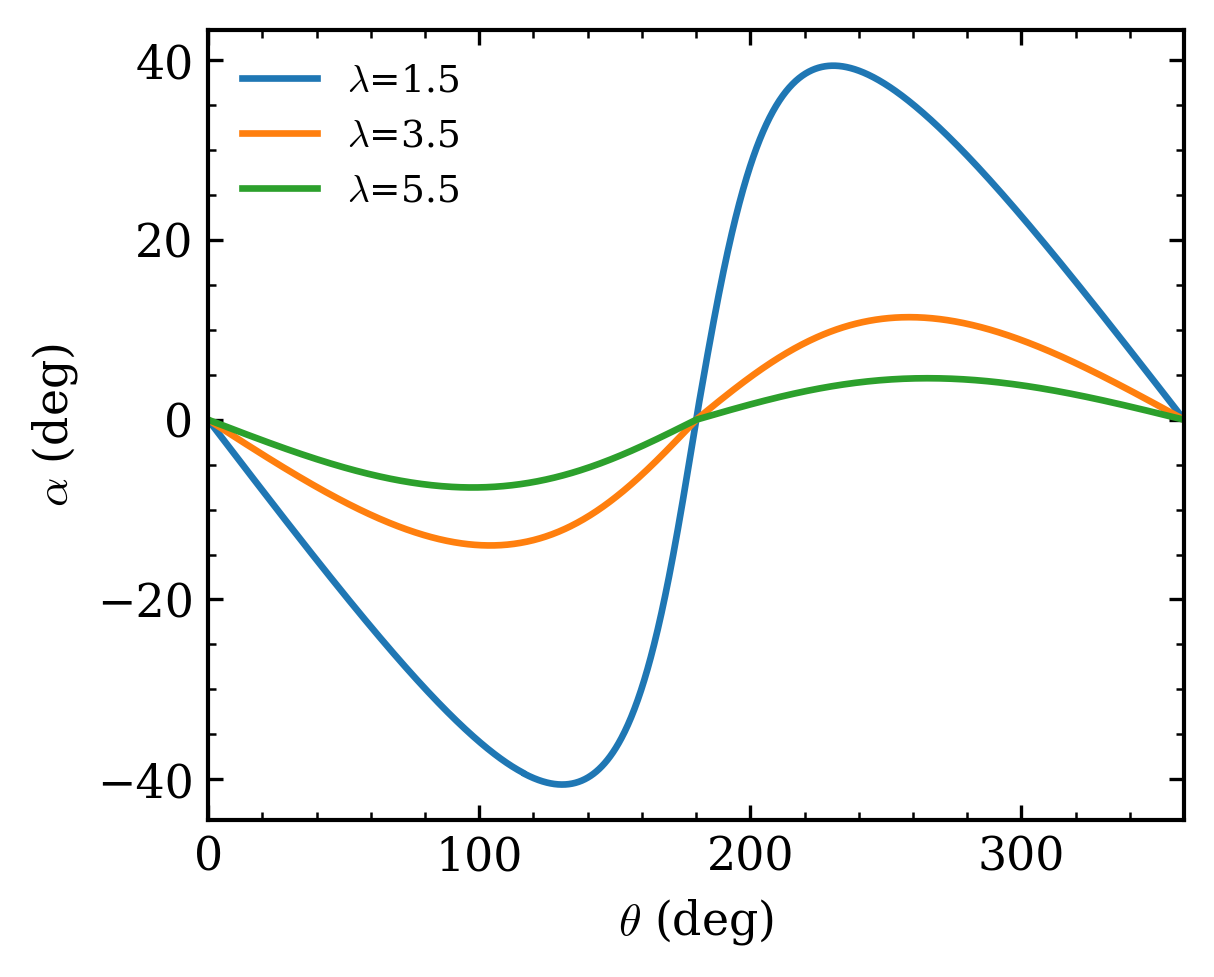}
        \\ (a) Angle of attack, $\alpha$
    \end{minipage}
    \hfill
    \begin{minipage}{0.48\linewidth}
        \centering
        \includegraphics[width=\linewidth]{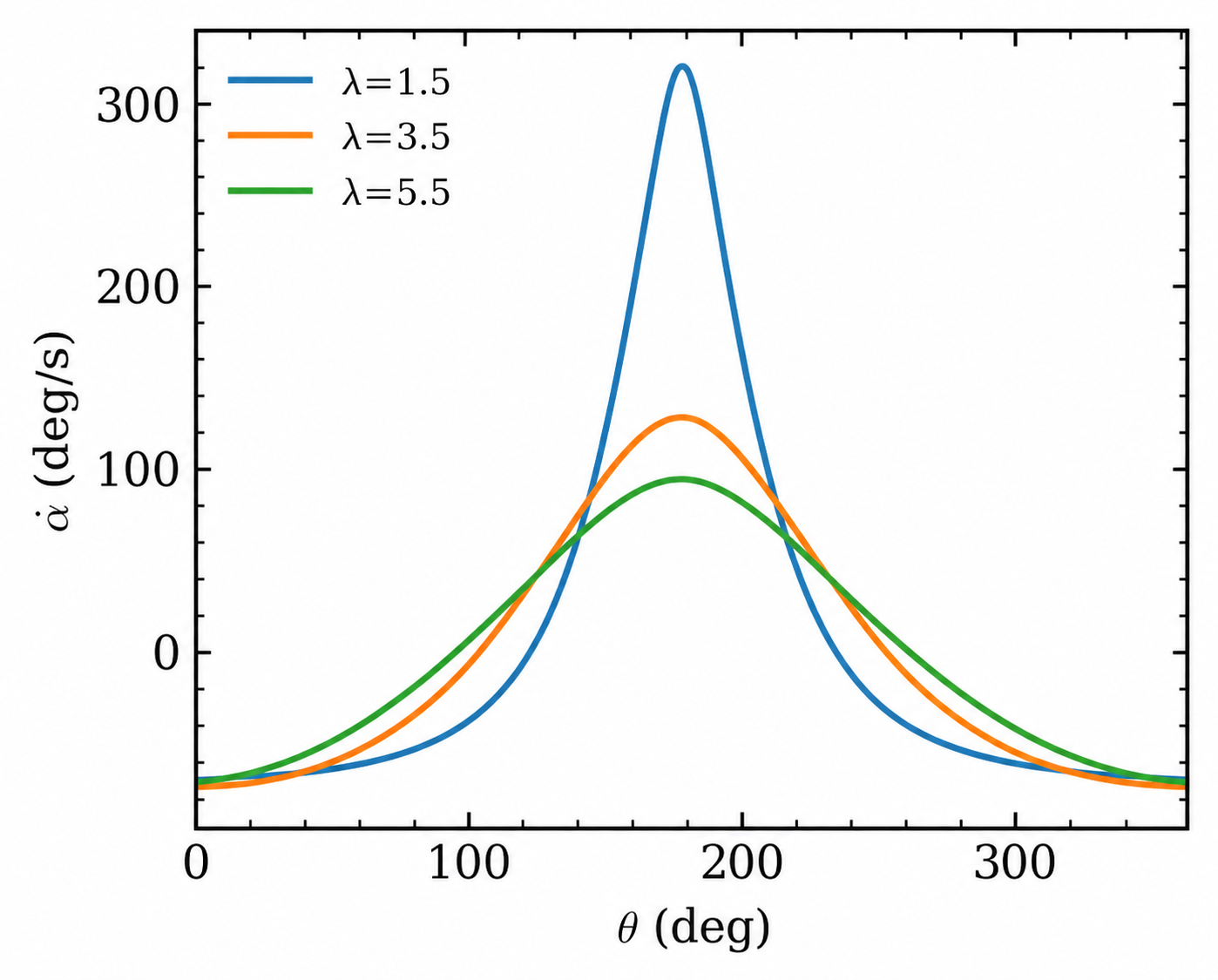}
        \\ (b) Rate of change, $\dot{\alpha}$
    \end{minipage}
    \caption{Phase-averaged (a) angle of attack $\alpha$ and (b) its rate of change $\dot{\alpha}$ as functions of azimuth $\theta$ for $\lambda=1.5$, $3.5$, and $5.5$.}
    \label{fig:alpha_alphadot_theta}
\end{figure}

Figure~\ref{fig:alpha_alphadot_theta} shows the phase-averaged angle of attack and its rate of change as functions of rotor azimuth for $\lambda=1.5$, $\lambda=3.5$, and $\lambda=5.5$. The amplitude of $\alpha$ decreases dramatically with increasing tip-speed ratio. At $\lambda=1.5$, the blade sweeps from approximately $-41^\circ$ in the upwind pass to $+40^\circ$ in the downwind pass, which places it deep into stall on both halves of the revolution. At $\lambda=3.5$, the excursion is confined to roughly $\pm12^\circ$ near the static stall boundary. At $\lambda=5.5$, the angle of attack remains approximately between $-8^\circ$ and $+5^\circ$, within the predominantly attached-flow regime. This reduction follows directly from the velocity triangle, in which the rotational contribution $\Omega R$ grows relative to the freestream as $\lambda$ increases and decreases the variation of incidence.

The rate of change $\dot\alpha$ in Fig.~\ref{fig:alpha_alphadot_theta}b peaks near
$\theta\approx180^\circ$, where the blade transitions between the upwind and downwind passes. The peak rate reaches approximately $320^\circ/\mathrm{s}$ at $\lambda=1.5$, more than double the values at $\lambda=3.5$ and $5.5$. This trend is consistent with the larger and more rapid incidence variation that drives the strong dynamic stall at low tip-speed ratio. The high incidence rate reached during the first half of the revolution at $\lambda=1.5$ activates the delayed separation during the upwind pass, whose effect on the blade loading is analyzed in Sec.~\ref{sec:ft_dynstall}.

\subsubsection{Instantaneous velocity and vorticity}
\label{sec:instantaneous}
\begin{figure}[h]
    \centering
    \begin{minipage}{0.48\linewidth}
        \centering
        \includegraphics[width=\linewidth]{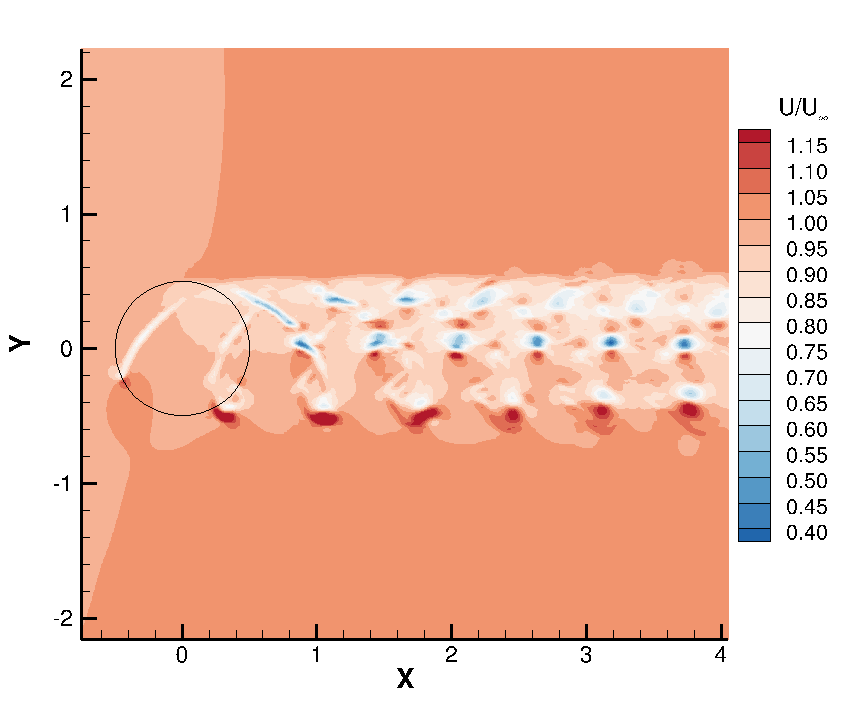}
        \\ (a) $\lambda=1.5$
    \end{minipage}
    \hfill
    \begin{minipage}{0.48\linewidth}
        \centering
        \includegraphics[width=\linewidth]{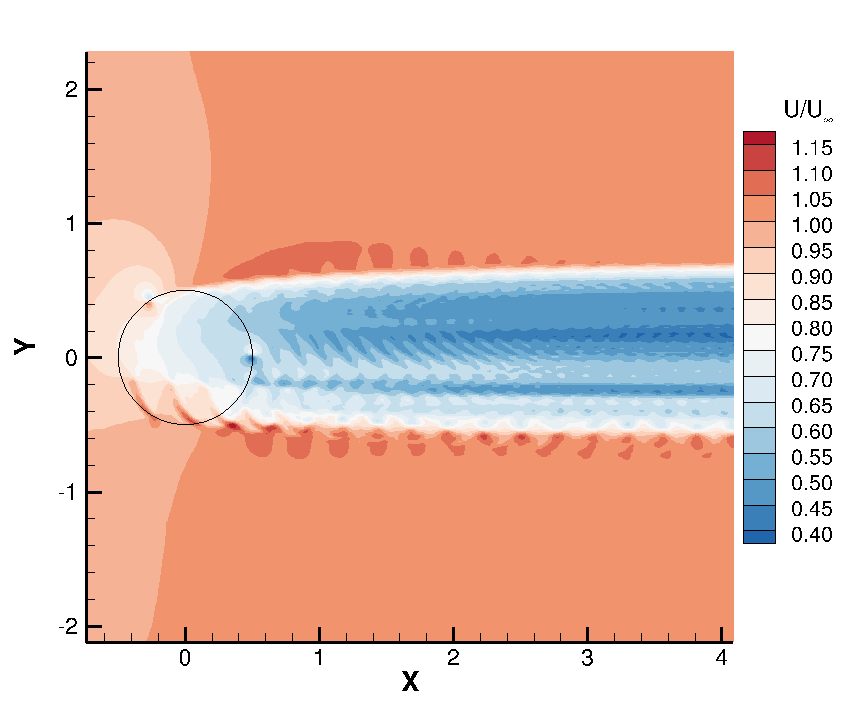}
        \\ (b) $\lambda=3.5$
    \end{minipage}
    \caption{Normalized instantaneous streamwise velocity contours at $t=20\,\mathrm{s}$ for (a)~$\lambda=1.5$ and (b)~$\lambda=3.5$.}
    \label{fig:inst_u}
\end{figure}

\begin{figure}[h]
    \centering
    \begin{minipage}{0.48\linewidth}
        \centering
        \includegraphics[width=\linewidth]{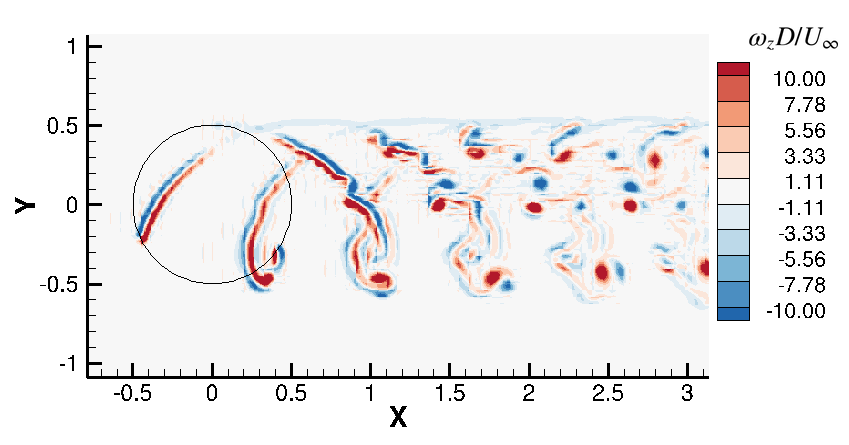}
        \\ (a) $\lambda=1.5$
    \end{minipage}
    \hfill
    \begin{minipage}{0.48\linewidth}
        \centering
        \includegraphics[width=\linewidth]{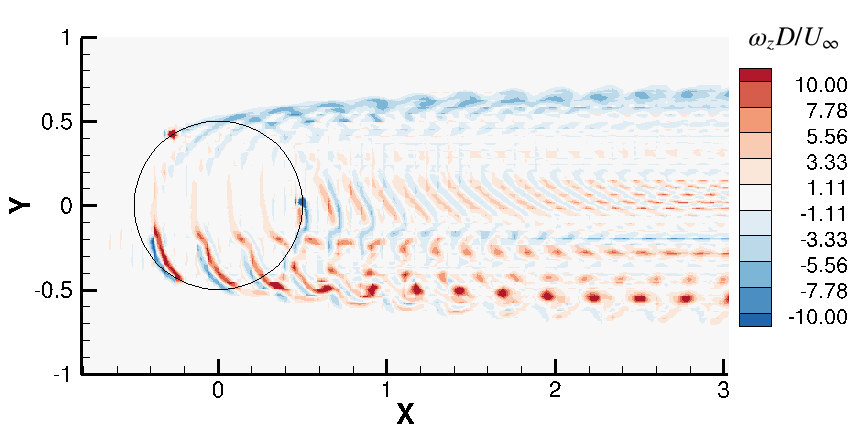}
        \\ (b) $\lambda=3.5$
    \end{minipage}
    \caption{Normalized instantaneous spanwise vorticity contours at $t=20\,\mathrm{s}$ for (a)~$\lambda=1.5$ and (b)~$\lambda=3.5$.}
    \label{fig:inst_vort}
\end{figure}

 In this subsection, only the flow fields from the $\lambda=1.5$ and $3.5$ cases are presented for comparison. Figures~\ref{fig:inst_u} and~\ref{fig:inst_vort} show the instantaneous streamwise velocity and spanwise vorticity at $t=20\,\mathrm{s}$ for $\lambda=1.5$ and $3.5$. The two operating conditions produce qualitatively different wakes. At $\lambda=1.5$, the modeled deep-stall loading generates large discrete vortical structures during successive blade passages. These structures spread over a broad lateral extent, with regions of freestream flow penetrating between them. At $\lambda=3.5$, the vorticity field is finer and more regularly organized into two shear layers surrounding a deeper central velocity deficit. Thus, the condition with stronger instantaneous vortex activity is not the condition with the larger mean velocity deficit, as examined in Sec.~\ref{sec:timeavg}.

\begin{figure}[h]
\centering
\begin{minipage}{0.48\linewidth}
    \centering
    \includegraphics[width=\linewidth]{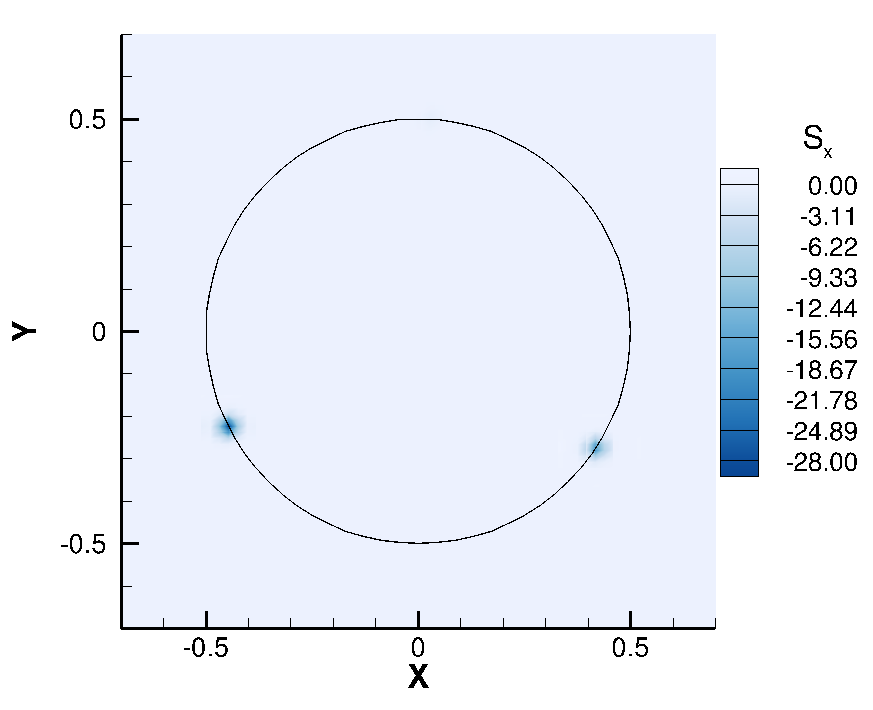}
    \\ (a) $S_x$
\end{minipage}
\hfill
\begin{minipage}{0.48\linewidth}
    \centering
    \includegraphics[width=\linewidth]{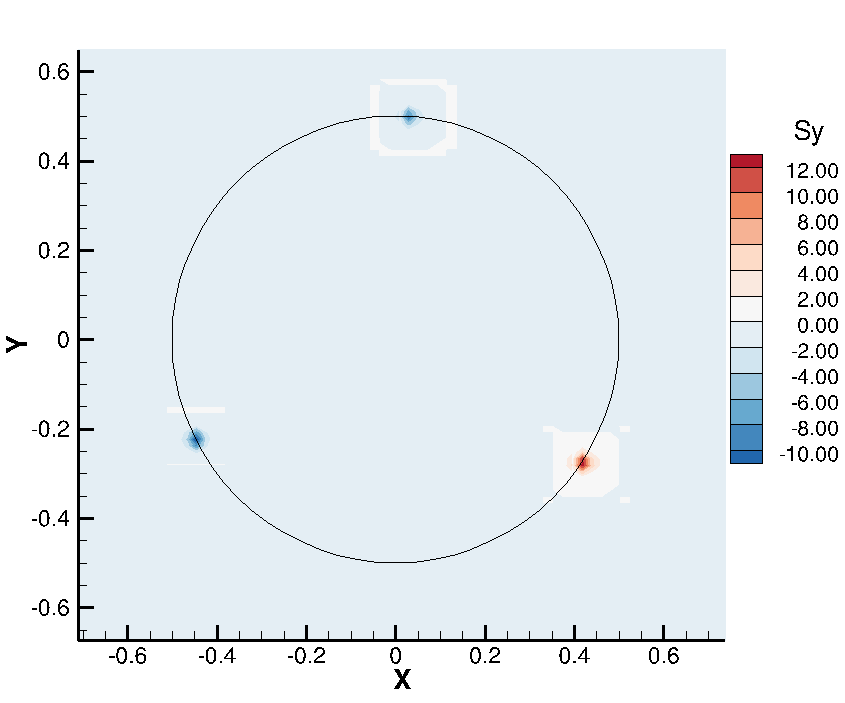}
    \\ (b) $S_y$
\end{minipage}
\caption{Instantaneous actuator-line momentum source terms $S_x$ and $S_y$ at $t=20\,\mathrm{s}$ for $\lambda=1.5$.}
\label{fig:alm_source_terms}
\end{figure}

Figure~\ref{fig:alm_source_terms} shows the corresponding momentum source terms $S_x$ and $S_y$ at $\lambda=1.5$. The compact regions of positive and negative forcing follow the instantaneous blade positions, and their smooth distribution around each blade position results from the Gaussian force projection. These distributed source terms produce the localized velocity and vorticity disturbances observed near the rotor in Figs.~\ref{fig:inst_u} and \ref{fig:inst_vort}.

\subsubsection{Time-averaged wake}
\label{sec:timeavg}
\begin{figure}[h]
    \centering
    \begin{minipage}{0.48\linewidth}
        \centering
        \includegraphics[width=\linewidth]{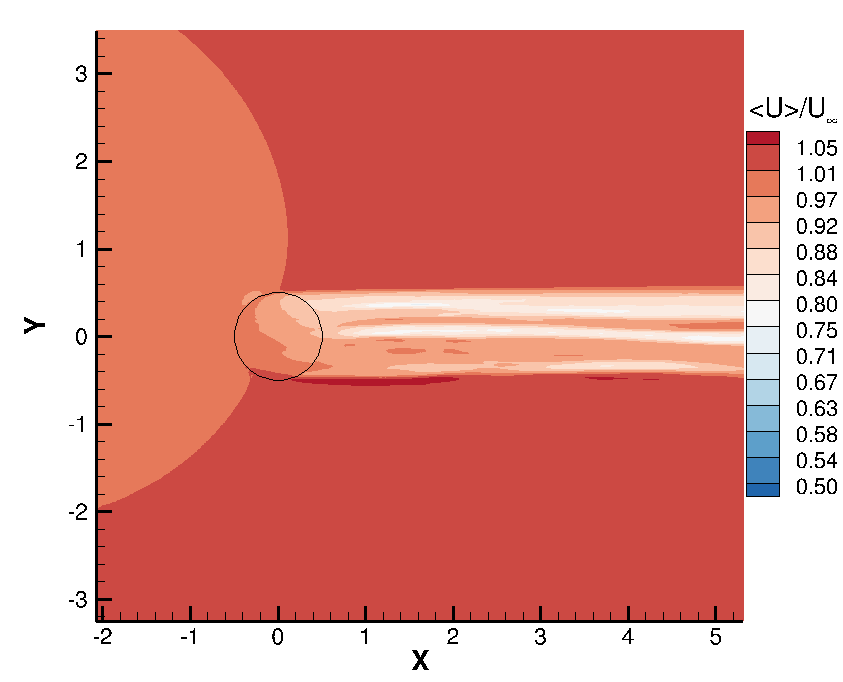}
        \\ (a) $\lambda=1.5$
    \end{minipage}
    \hfill
    \begin{minipage}{0.48\linewidth}
        \centering
        \includegraphics[width=\linewidth]{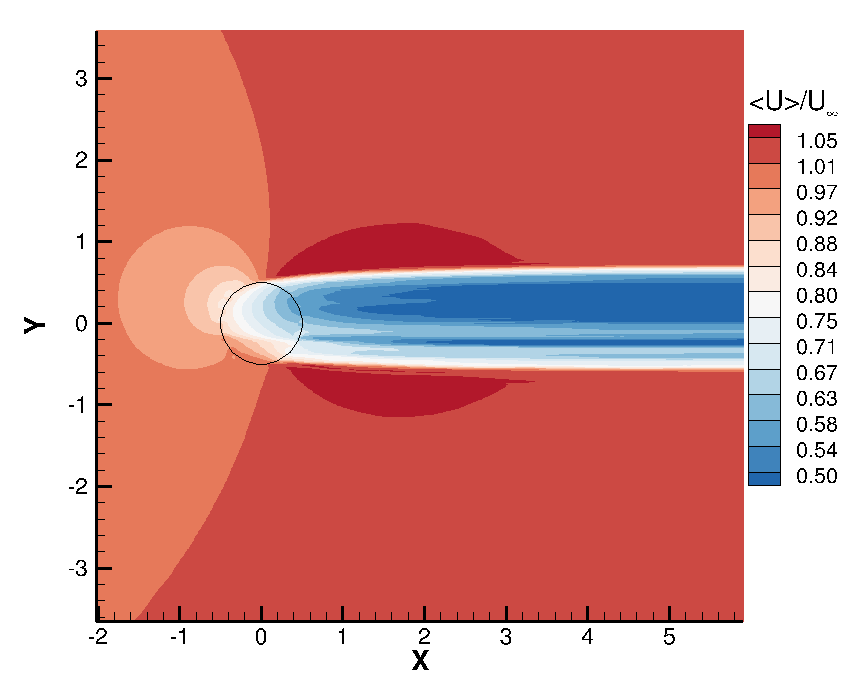}
        \\ (b) $\lambda=3.5$
    \end{minipage}
    \caption{ Normalized time-averaged streamwise velocity contours for (a)~$\lambda=1.5$ and (b)~$\lambda=3.5$.}
    \label{fig:TimeAvgd_u}
\end{figure}

\begin{figure}[h]
    \centering
    \begin{minipage}{0.48\linewidth}
        \centering
        \includegraphics[width=\linewidth]{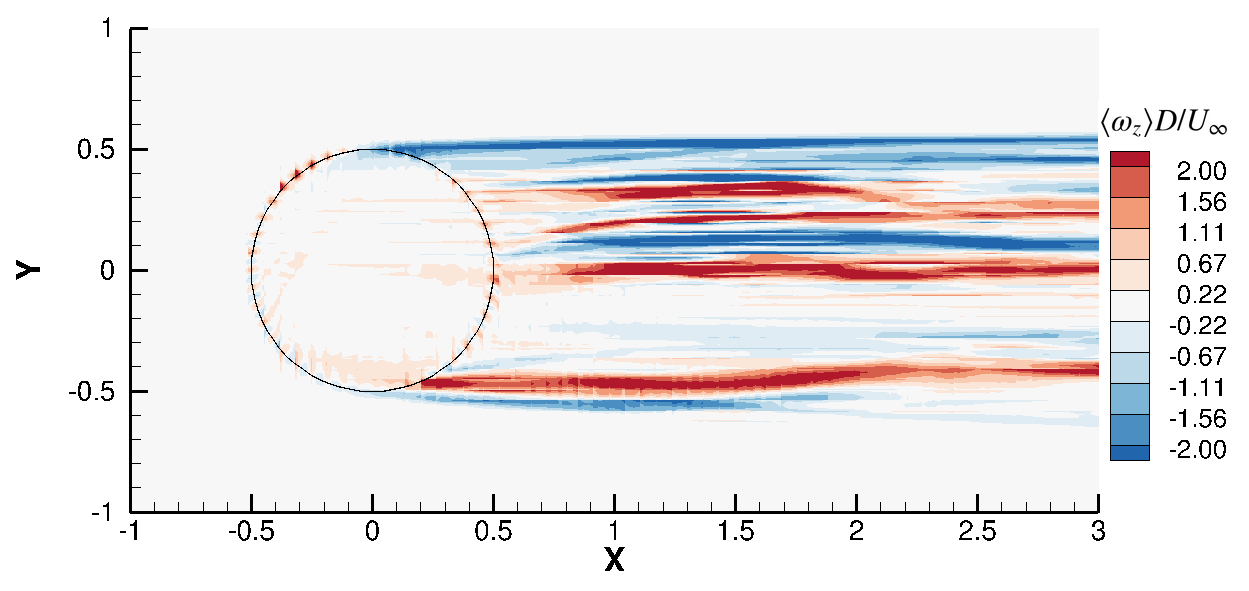}
        \\ (a) $\lambda=1.5$
    \end{minipage}
    \hfill
    \begin{minipage}{0.48\linewidth}
        \centering
        \includegraphics[width=\linewidth]{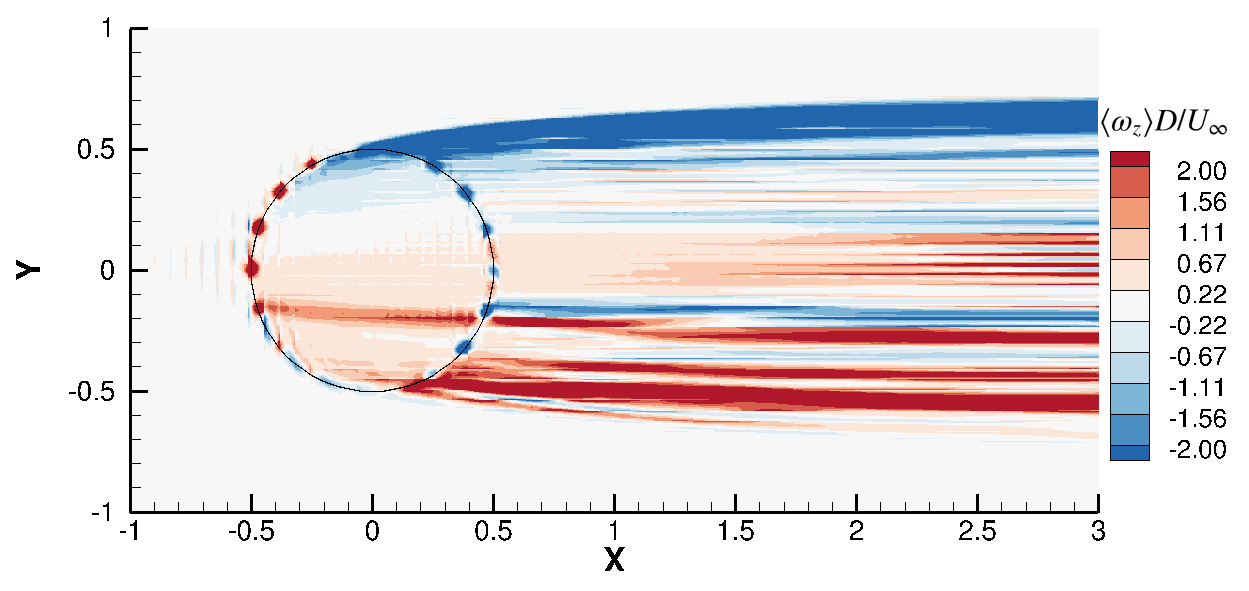}
        \\ (b) $\lambda=3.5$
    \end{minipage}
    \caption{Normalized time-averaged spanwise vorticity contours for (a)~$\lambda=1.5$ and (b)~$\lambda=3.5$.}
    \label{fig:TimeAvgd_vorticity}
\end{figure}

Figures~\ref{fig:TimeAvgd_u} and~\ref{fig:TimeAvgd_vorticity} show the time-averaged streamwise velocity and spanwise vorticity fields. At $\lambda=1.5$, large discrete vortical structures are prominent in the instantaneous field, whereas the mean wake is broad and its central velocity deficit is comparatively shallow. This condition also has a low power coefficient, $C_P\approx0.03$. At $\lambda=3.5$, the instantaneous wake is more regularly organized, while the mean velocity deficit is deeper and more elongated and the power coefficient is substantially higher, $C_P\approx0.38$. The mean vorticity is concentrated within two persistent shear layers that bound the wake. For the two conditions examined, the intensity of instantaneous vortex activity and the depth of the mean velocity deficit therefore exhibit different trends with tip-speed ratio.

The asymmetric mean deficit and unequal development of the wake shear layers are consistent with the experimental and numerical wake structures reported by Tescione et al.~\cite{tescione2014near} and Posa et al.~\cite{posa2016wake}. Araya et al.~\cite{araya2017transition} further showed that the downstream development of a VAWT wake depends on the operating condition and evolves from blade-scale periodic shedding toward bluff-body-like wake dynamics. The present results complement these studies by relating the operating-regime-dependent instantaneous wake structure to the corresponding mean velocity deficit and turbine power.

\subsection{Computational efficiency}
\label{sec:efficiency}
The representative case at $\lambda=3.5$ requires about $37$ minutes of wall-clock time on $48$ CPU cores (or roughly one minute per revolution) to advance the flow field to $30\,\mathrm{s}$ of physical time ($33.4$ rotor revolutions). This rapid turnaround makes parametric tip-speed-ratio sweeps practical. This efficiency is accompanied by reduced near-wall fidelity, since body-fitted methods resolve the blade boundary layer directly and predict drag-sensitive quantities such as separation onset and total pressure loss more accurately than the present approach. The framework is therefore suited to efficient comparative assessment of sectional blade loading, overall rotor performance, and near-wake development across VAWT operating conditions.

\section{Conclusion}
\label{sec:conclusion}
This study examined how dynamic stall modeling alters the distribution of VAWT blade loading between the upwind and downwind passages and how instantaneous wake structure changes relative to the mean velocity deficit across operating regimes of a VAWT. At $\lambda=1.5$ and $3.5$, the influence of the dynamic stall correction arose predominantly during the upwind passage. At $\lambda=1.5$, the correction increased the mean upwind tangential force by nearly a factor of three and increased the power coefficient by $23\%$. This relative increase was small in absolute terms, however, and the resulting power coefficient remained only about $0.03$, confirming that the transient loading enhancement under deep dynamic stall did not translate into favorable overall turbine performance. At $\lambda=3.5$, the correction reduced the mean upwind tangential force by approximately $23\%$ and decreased the power coefficient by $11\%$. The downwind contribution changed comparatively little at both conditions. At $\lambda=5.5$, the two loading histories differed only modestly, with little change in the mean upwind loading and a small increase in the downwind contribution. The corresponding power coefficient increased by $7\%$. These results show that the modeled dynamic stall contribution is not a uniform adjustment to turbine performance but depends on the operating regime and the phase of the blade motion.

The wake comparison between $\lambda=1.5$ and $3.5$ revealed a second distinction between instantaneous and mean flow behavior. The deep stall condition at $\lambda=1.5$ produced larger and more widely distributed instantaneous vortical structures, but its mean wake had a comparatively shallow velocity deficit and its power coefficient was only about $0.03$. At $\lambda=3.5$, the instantaneous wake was more regularly organized, whereas the mean velocity deficit was deeper and the power coefficient increased to approximately $0.38$. Thus, for the two conditions investigated, stronger instantaneous vortex activity did not correspond to a deeper mean wake or greater power extraction. This contrast highlights the need to distinguish phase-dependent unsteadiness from time-averaged turbine performance when interpreting VAWT wake behavior.

The analysis was enabled by a rotating FR/CPR-ALM on fixed Cartesian grids, incorporating Gaussian force projection, a modified Boeing--Vertol dynamic stall model, and a double-multiple-streamtube-based induction closure. The predicted mean near-wake profile reproduced the principal deficit and asymmetry observed in experimental and LES-ALM reference data, while the power-coefficient curve followed the three-dimensional LES-ALM reference near the optimal operating range, with a peak difference of approximately $6\%$. The Gaussian-width analysis further distinguished mesh-controlled and chord-controlled projection regimes and clarified the relationship between the force-projection width and spatial resolution. The framework therefore provides an efficient approach for comparing phase-dependent loading and wake development over multiple operating conditions, while its two-dimensional formulation and reliance on tabulated aerodynamic data limit its representation of spanwise transport, finite-blade effects, and blade-boundary-layer physics. Future work should assess these findings in three-dimensional configurations and across additional turbine geometries and inflow conditions.

\section*{Acknowledgments}
Part of the hardware used in the computational studies is from the UMBC High Performance Computing Facility (HPCF). The facility is supported by the U.S. National Science Foundation through the MRI program (grant nos. CNS-0821258, CNS-1228778, OAC-1726023, and CNS-1920079) and the SCREMS program (grant no. DMS-0821311), with additional substantial support from the University of Maryland, Baltimore County (UMBC).

\bibliography{aipsamp}

\end{document}